\documentclass{nature}

\usepackage{graphicx}
\usepackage{color}
\usepackage{float}
\usepackage{soul} 
\usepackage{textcomp}

\def\UHH{$U_\mathrm{HH}$}  
 
\def\Ef{$E_F$}
\def\oC{$^{\circ}$C}

\newcommand{\li}[1]{{\color{black} #1}}
\newcommand{\np}[1]{{\color{black} #1}}

\setstcolor{red}

\title{Single Spin Localization and Manipulation \\ in Graphene Open-Shell Nanostructures}

\author
{Jingcheng Li$^{1}$, Sofia Sanz$^{2}$, Martina Corso$^{2,3}$, Deung Jang Choi$^{2,3,5}$,  Diego Pe{\~{n}}a$^{4}$, \\ 
Thomas Frederiksen$^{2,5}$, Jose Ignacio Pascual$^{1,5}$ \\
\\
\normalsize{$^{1}$ CIC nanoGUNE, 20018 Donostia-San Sebasti\'an, Spain}\\
\normalsize{$^{2}$Donostia International Physics Center (DIPC), 20018 Donostia-San Sebasti\'an, Spain}\\
\normalsize{$^{3}$ Centro de F{\'{\i}}sica de Materiales 	CFM/MPC (CSIC-UPV/EHU),  20018 Donostia-San Sebasti\'an, Spain}\\
\normalsize{$^{4}$Centro Singular de Investigaci\'on en Qu\'imica Biol\'oxica e Materiais Moleculares (CiQUS),} \\
\normalsize{and Departamento de Qu\'imica Org\'anica, Universidade de Santiago de Compostela, Spain}\\
\normalsize{$^{5}$Ikerbasque, Basque Foundation for Science, 48013 Bilbao, Spain}
\\
\\
}

\makeatletter

\let\saved@includegraphics\includegraphics
\AtBeginDocument{\let\includegraphics\saved@includegraphics}
\renewenvironment*{figure}{\@float{figure}}{\end@float}
\makeatother

\begin{document}

\maketitle
\date{\today}

\begin{abstract}
Predictions state that graphene can spontaneously develop magnetism from the Coulomb repulsion of its $\pi$-electrons, but its experimental verification  has been  a challenge. Here, we report on the observation and manipulation of individual magnetic moments localized in graphene nanostructures on a Au(111) surface. Using scanning tunneling spectroscopy, we detected the presence of single electron spins localized around certain zigzag sites of the carbon backbone via the Kondo effect. Two near-by spins were found coupled into a singlet ground state, and the strength of their exchange interaction was measured via singlet-triplet inelastic tunnel electron excitations.  Theoretical simulations demonstrate that electron correlations result in spin-polarized radical states with the experimentally observed spatial distributions. Hydrogen atoms bound to these radical sites quench their magnetic moment, permitting  us to  switch the spin of the nanostructure using the tip of the microscope.
\end{abstract}
\linespread{1.2}

\newpage
Among the many applications predicted for graphene, its use as a source of magnetism is the most unexpected one,
and an attractive challenge for its active role in spintronic devices \cite{Han2014}. 
Generally, magnetism is associated to a large degree of electron localization and strong spin-orbit interaction. Both premises are absent in graphene, a strongly diamagnetic material. 
The simplest method to induce magnetism in graphene is to create an imbalance in the number of  carbon atoms in each of the two sublattices, what, according to the Lieb's theorem for bipartite lattices \cite{Lieb1989}, causes a spin imbalance in the system. This can be done by either inserting defects that remove a single ${p_z}$ orbital \cite{Yazyev2007,Nair2012,McCreary2012,Gonzalez2016} or by shaping graphene with zigzag edges \cite{Fernandez-Rossier2007,Yazyev2010}.
However, magnetism can also emerge  in graphene nanostructures where  Lieb's theorem does not apply \cite{Alexandre2012,Ortiz2016}. For example, in $\pi$-conjugated systems with small band gaps, Coulomb repulsion between valence electrons forces the electronic system to reorganize into open-shell configurations \cite{Morita2011} with unpaired electrons (radicals) localized at different atomic sites. Although the net magnetization of the nanostructures may be zero, each radical state hosts a  magnetic moment of size $\mu_B$, the Bohr magneton, turning the graphene nanostructure  paramagnetic. This basic  principle predicts, for example, the emergence of
 edge magnetization originating from zero-energy modes in sufficiently wide zigzag \cite{Son2006,Tao2011,Ruffieux2016} and chiral \cite{Yazyev2011,Carvalho2014} graphene nanoribbons (GNRs). 

The  experimental observation of  spontaneous magnetization driven by electron  correlations is still challenging, because, for example, atomic defects and impurities in the graphene structures \cite{Sepioni2010,Nair2013} hide the weak paramagnetism of radical sites \cite{Cervenka2009}. Scanning probe microscopies can spatially localize the source states of magnetism, but they require  both atomic-scale resolution and spin-sensitive measurements.   Here we achieve these conditions to demonstrate that atomically-defined  graphene nanostructures  can host localized spins at specific sites and give rise to the Kondo effect \cite{Kondo1964,Ternes2009}, a many-body phenomenon caused by the interaction between a  localized spin and free conduction electrons in its proximity.
Using a low-temperature scanning tunneling microscope (STM) we  use this signal to  map the spin localization within the nanostructure and to detect spin-spin interactions.

\begin{figure}
\includegraphics[width=\textwidth]{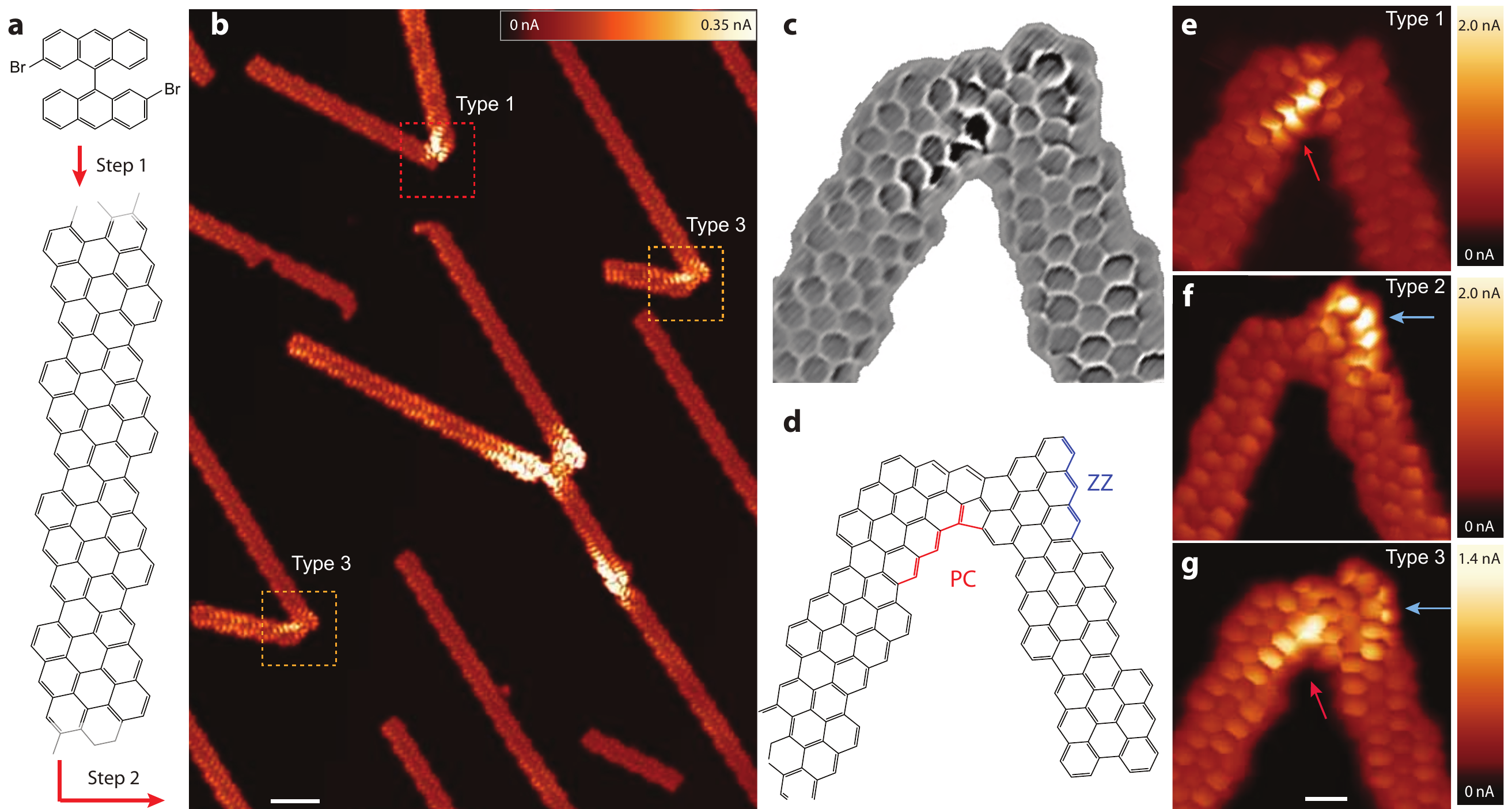}
\caption{\textbf{Formation of graphene junctions by cross-dehydrogenative coupling of adjacent graphene 
nanoribbons}. \textbf{a}, Model structures of the organic precusor 2,2'-dibromo-9,9'-bianthracene and of the on-surface synthesized (3,1)chGNR  after Ullmann-like C-C coupling reaction and 
cyclodehydrogenation on Au(111). 
\textbf{b}, Constant-height current images  ($V=2$ mV, scale bar: 2 nm) showing joint chGNR nanostructures,  \ with an angle of $\sim$50$^\circ$, 
obtained  after further annealing the sample.   A CO-functionalized tip was used to resolve the chGNR ring structure. Dashed boxes indicate  the most characteristic chGNR junctions, whose structure is shown in panels \textbf{c,d}.  
\textbf{c},  Laplace-filtered image of the junction shown in panel g to enhance the backbone structure, and \textbf{d}, model structure of the junction. PC labels the pentagonal cove site and the ZZ the zigzag site. 
\textbf{e-g}, Constant-height current images ($V=8$ mV, scale bar 0.5 nm) of the three types of chGNR junctions with same backbone structures but  with different LDOS distribution. 
}
\end{figure}

The graphene nanostructures studied here are directly created  
on a Au(111) surface  by cross-dehydrogenative coupling of adjacent chiral GNRs (chGNRs)\cite{Dienel2015}. We deposited  the organic molecular precursors  2,2'-dibromo-9,9'-bianthracene (Fig.~1\textbf{a}) on a clean Au(111) surface, 
and  annealed stepwisely to 250~\oC\ (step 1 in Fig.~1\textbf{a}) to produce narrow chGNRs\cite{Ruffieux2016,Dimas2016}. 
The chiral ribbons are semiconductors with a band-gap of 0.7 eV (Supplementary Figs. S7 and S8) and show two enantiomeric forms on the surface\cite{Merino2017}. 
By further annealing the substrate to 350 
\oC\ (step 2 in Fig.~1\textbf{a}),  chGNRs  fuse together into junctions, as shown in  Fig.~1\textbf{b}.  The chGNR junctions  highlighted by  dashed rectangles  are the most frequently found in our experiments. They consist of two chGNRs with the same chirality 
linked together by their termination (Fig.~1\textbf{c}).  The creation of this stable nanostructure implies the reorganization of the carbon atoms around the 
initial contact point\cite{Shiotari2017} into the final structure shown in Fig.~1\textbf{d},   \li{ as described in Supplementary Fig.~S1}.

In  Fig.~1\textbf{b}, certain regions of the junctions appear brighter when recorded at low sample bias, reflecting enhancements of the local density of states (LDOS) close to the Fermi level.
Interestingly, the precise location of the bright regions is not unique, but can be 
localized over the pentagon cove (PC) site (Type 1, Fig.~1\textbf{e}), over the terminal zigzag 
(ZZ) site of the junction (Type 2, Fig.~1\textbf{f}), or over both (Type 3, Fig.~1\textbf{g}). Despite these 
different LDOS distributions in the three types of junctions, they have identical
carbon   arrangement (Fig.~1\textbf{d}).

To understand the origin of the enhanced LDOS at the ZZ and PC sites, we recorded differential conductance spectra ($dI/dV$) on the three types of 
junctions. Spectra  on the bright sites of Type 1 and 2 junctions  show 
very pronounced zero-bias peaks   (Fig.~2\textbf{a,b}) localized over the  bright sites (spectra 1 to 4, and 6 to 8), and vanishing rapidly in neighbor rings  (spectra 5, 9, and 10). These are generally ascribed as Abrikosov-Suhl resonances due to the Kondo 
effect, and named as Kondo resonances\cite{Kondo1964,Ternes2009}.
Their observation  is a proof of  a localized 
magnetic moment  screened by conduction electrons\cite{Fernandez-Torrente2008,Zhang2013}.
The resonance line width increases with temperature (Fig.~2\textbf{d}) and magnetic field (Fig.~2\textbf{e}) following the characteristic behavior of a spin-\textonehalf\ system  with a  Kondo 
Temperature  $T_K \sim 6$ K\cite{Nagaoka2002,Zhang2013}. 

\begin{figure}
\includegraphics[width=0.95\textwidth]{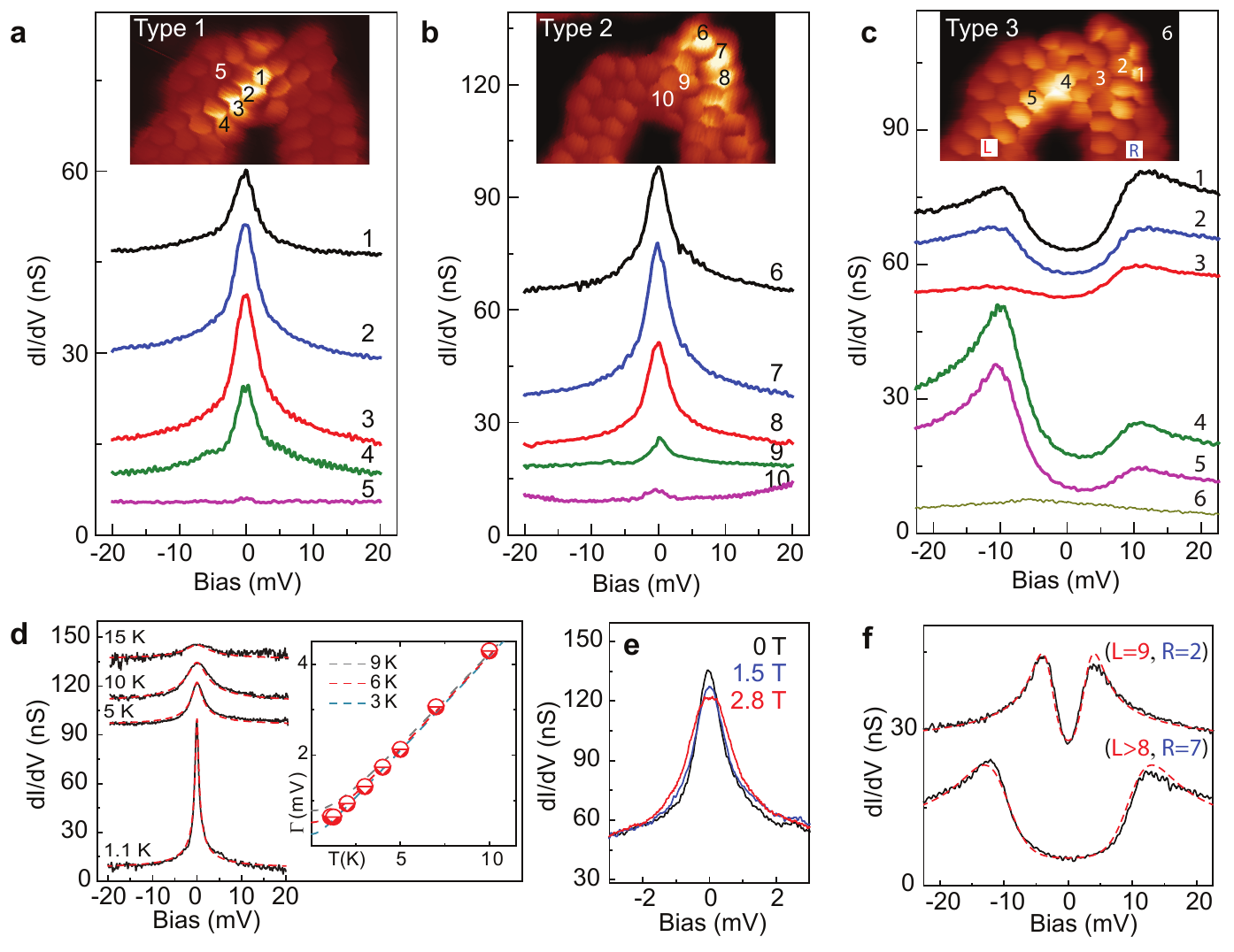}
\caption{\textbf{Spatial distribution of Kondo resonances and singlet-triplet excitations in chGNR junctions}. \textbf{a,b}, Kondo resonances  over the bright regions of 
Type 1 and Type 2 junctions, respectively.
The zero-bias peaks are mostly detected over four PC rings of 
Type 1 junctions and over three ZZ rings of Type 2 junctions.
\textbf{c}, Double-peak features around zero bias over Type 3 junctions. 
\textbf{d}, Temperature dependence of the Kondo resonance. All spectra were measured over the same PC site.  
The full width at half maximum (FWHM) at each temperature is extracted by fitting a Frota function (red dashed lines)\cite{Frota1992}, and corrected for the thermal broadening of the tip\cite{Zhang2013}. The temperature dependence of FWHM was fit by the empirical expression  $\sqrt{(\alpha k_BT)^2+(2k_BT_K)^2}$\cite{Nagaoka2002}, resulting in a Kondo temperature $T_K \sim$ 6 K and $\alpha=9.5$. 
\textbf{e}, Magnetic field dependence of a  Kondo resonance (over the same PC site) at the field strengths indicated in the figure. 
\textbf{f},  Split-peak $dI/dV$ features for nanostructures  with different sizes, determined by the number of precursor units in each chGNR, labeled \textsf{L} and \textsf{R} in panel \textbf{c}. 
The gap width 
increases with the length of the ribbons \li{(Supplementary Section 7)}. The red dashed lines are fits to our spectra  using a model for two coupled spin-\textonehalf\ systems\cite{Ternes2015}. The spectra in \textbf{d} and \textbf{e} were acquired with a metal tip, while the others with a CO-terminated tip. }
\end{figure}

Junctions with two bright regions (Type 3) show   
different low-energy features: two peaked steps in $dI/dV$ spectra at $\sim \pm$10 meV 
(Fig.~2\textbf{c}). The peaks appear at the same energies over the terminal ZZ segment and over 
the PC region for a given nanostructure, and  vanish quickly away from these sites. Based on the 
existence of localized spins on  bright areas of Type 1 and 2 junctions, we attribute the 
double-peak features to excitation of two exchange coupled spins  localized  at each 
junction site. 
The exchange interaction tends to freeze their relative orientation, in this case
antiferromagnetically into a singlet ground state. Electrons tunneling into the coupled 
spin system can excite a spin reversal in any of them when their energy equals the exchange coupling energy between the spins, i.e., $eV\geq J$.  
Usually, this inelastic process is revealed in $dI/dV$ spectra as   
steps at the onset of spin excitations\cite{Hirjibehedin2006}, from which one can directly determine the strength of the exchange coupling $J$ between the spins.  
Here, the spectra additionally show asymmetric peaks on top of the excitation onsets
characteristic of Kondo-like systems with particle-hole asymmetry, when spin fluctuations are  hindered in the ground state\cite{Heersche2006,Paaske2006,Ternes2015,Ortiz2017}. 
Hence, the gap between $dI/dV$ peaks in Fig.~2\textbf{c} is a measure of the interaction strength between the two localized spins. 

Interestingly, the spectral gap in Type 3 junctions increases with the length  of the connecting ribbons (See Supplementary Fig.~15b).
In Fig. 2\textbf{f} we  compare  low-energy spectra of two junctions with different chGNR lengths.
Although  the atomic structures of both junctions  are identical, the one  with shorter ribbons (upper curve; 9 and 2 precursor units) displays 
a smaller gap than the junction of longer chGNRs (lower curve; $>8$ and 7 units). Fitting the spectra with a model of two coupled spin-\textonehalf\ systems\cite{Ternes2015}, one obtains the exchange coupling $J=2.7$ ($9.9$) meV for the upper (lower) spectrum.

\begin{figure} 
	\includegraphics[width=\textwidth]{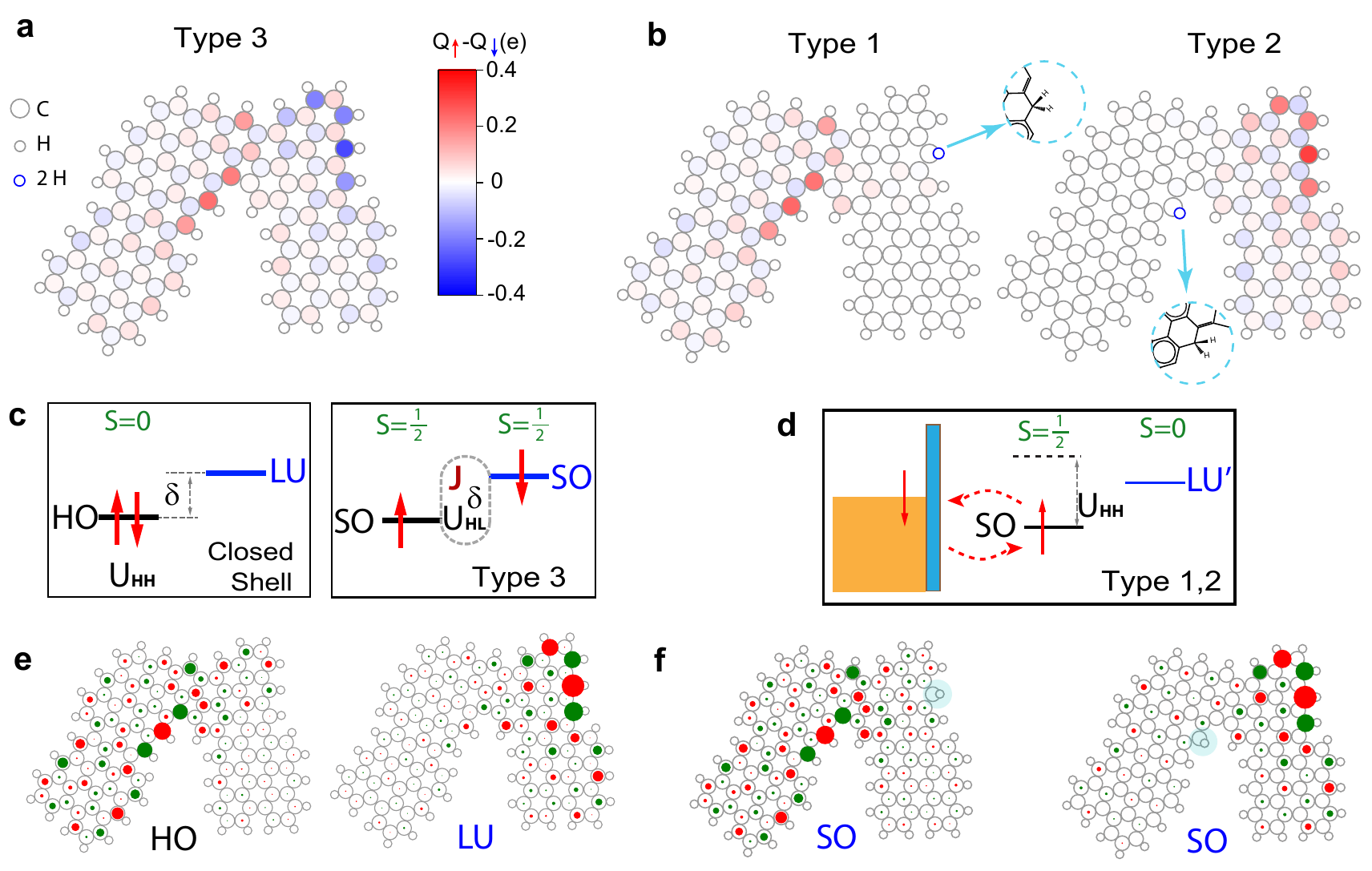}
	\caption{\textbf{Calculated electronic states of chGNR junctions}. \textbf{a,b}, Spin polarization obtained from DFT simulations in a Mulliken population analysis. 
    The standard junction shown in \textbf{a} (all peripheral carbons bonded to H) shows 
    spontaneous spin localization in both PC and ZZ regions, revealing the apparition of 
    radical states. Adding a H atom to an external carbon in either the ZZ (Type 1) or PC 
    (Type 2) removes the corresponding radical state and, hence, its  spin-polarization.
    	\textbf{c}, Schema of the spontaneous spin polarization when one of the two electrons
	in the HO level gets promoted to the LU level to form two separated, exchange coupled spin-\textonehalf\ systems (Type 3 junction). This process is energetically favored when the reduction in Coulomb energy $U_\mathrm{HH}-U_\mathrm{HL}$ plus exchange energy $J$ exceeds the level separation $\delta$, i.e., $\delta+U_\mathrm{HL}-J<U_\mathrm{HH}$.
	\textbf{d}, Sketch of the spin-\textonehalf\ Kondo state generated with a single radical (Type 1 and 2 junctions).
	\textbf{e}, Single-particle TB wave functions (HO/LU) for Type 3 junction.
    \textbf{f}, Single-particle TB wave functions (SO) for Type 1 and Type 2 junctions.
    Red-green colors represent the positive-negative phase. 
    }
\end{figure}

\np{To explain the emergence of localized spins}, we simulated the 
spin-polarized electronic structure of chGNR junctions using both
density functional theory  (DFT) 
and mean-field Hubbard (MFH) models (see Supplementary Sections 5 and 6).
Fig. 3\textbf{a} shows the spin-polarization of a junction of Fig.~1\textbf{d}. The ground state exhibits a net spin localization  at the ZZ and PC regions with opposite sign, \np{which is absent in the bare ribbons. This spin distribution agrees with the observations for Type 3 junctions. }
The origin of the spontaneous magnetization can be rationalized by considering  the effect 
of Coulomb correlations between $\pi$-electrons as described within a tight-binding (TB) model.
The spin distribution is related to the appearance of two junction states 
inside the gap of the (3,1)-chGNR electronic bands, localized at the PC and ZZ sites, respectively.
In the absence of electron-electron correlations, these two states conform the highest occupied (HO) and lowest unoccupied (LU) molecular states of the nanostructure (Fig.~3\textbf{e}).
Due to the large degree of localization 
(Supplementary Figs.~S10-S11), the Coulomb repulsion energy \UHH\ between two electrons in the HO state becomes comparable  with
the energy difference $\delta$ between the two localized levels. Hence, in a simplified picture, the two electrons find a lower-energy configuration by occupying 
each a different, spatially separated in-gap state.
These two states become singly occupied (SO), spin-polarized (i.e.,
they have a net magnetic moment), and exchange coupled as schematically illustrated in Fig.~3\textbf{c}. 
Similar  conclusions  regarding the emergence  of radical states at PC and ZZ sites can also be reached using the empirical Clar's aromatic $\pi$-sextet rule (Supplementary Section 3).

According to both DFT (Fig.~3\textbf{a}) and MFH (Supplementary Fig.~S9) the magnetic moments are 
antiferromagnetically aligned into a singlet ground state. 
Therefore, the inelastic features in $dI/dV$ spectra found over Type 3 junctions (Fig.~2\textbf{c}) are associated to singlet-triplet excitations induced by tunneling electrons. 
\np{In fact, the smaller excitation energy found for the smaller ribbons 
in both theory and experiment (Supplementary Section 7) agrees with a weaker exchange interaction due to a larger localization of the spin-polarized states. Alternative scenarios for peaks around \Ef, such as single-particle states or 
Coulomb-split radical states\cite{Gonzalez2016}, would show the opposite trend with
the system size.}

To account  for spin localization in only one of the two radical regions in Type 1 and 2 junctions,  one of the two edge magnetic moments has  to vanish.  H-passivation of radical sites is a highly probable process occurring on the  surface due to the large amount of hydrogen available during the reaction\cite{Talirz2013}. 
DFT simulations show that attaching an extra H atom into  an edge 
carbon in either the ZZ (Type 1) or PC (Type 2) sites leads to 
its $sp^3$ hybridization and  the removal of a $p_z$ orbital from the aromatic backbone. 
This completely quenches the magnetic moment of the passivated region, and leaves the junction 
with a single electron localized at the opposite radical site (Supplementary Fig.~S6). 
The computed distributions for the two energetically most favorable adsorption sites (Fig.~3\textbf{b}) are in excellent agreement with the extension of the Kondo resonance mapped in Fig.~2\textbf{a,b}. 

\begin{figure}[!htb]
\begin{center}
	\centering
	\includegraphics[width=0.7\columnwidth]{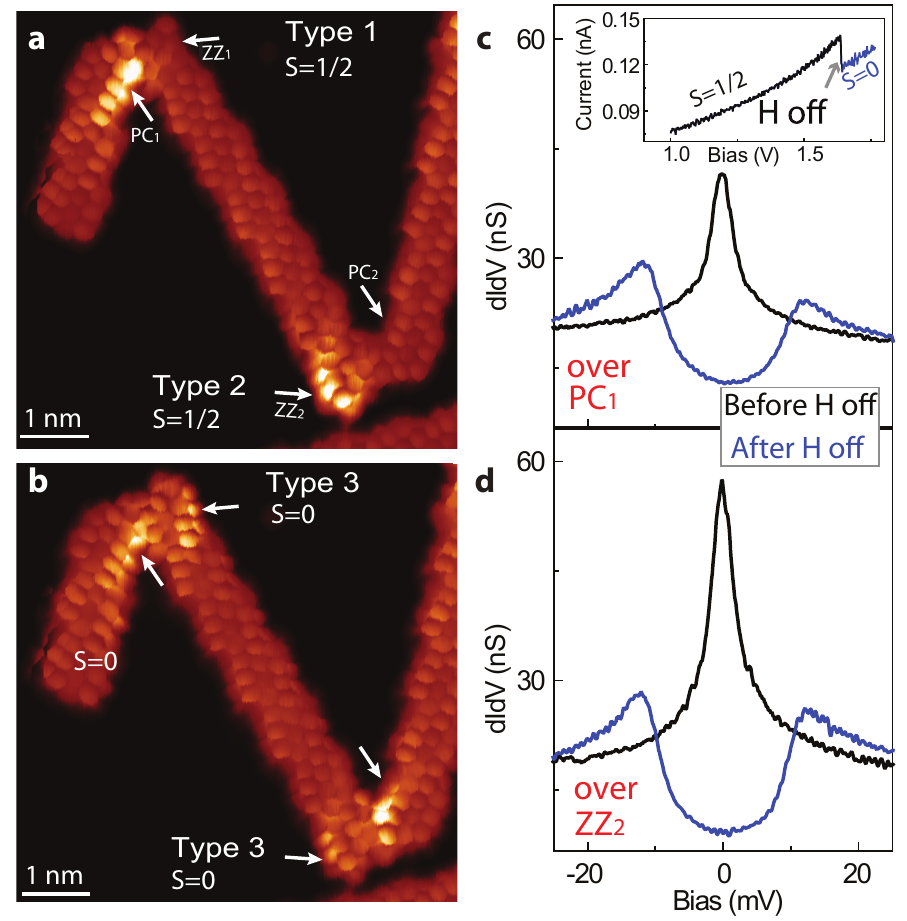}
	\caption{\textbf{Spin manipulation by electron-induced removal of extra H-atoms}. 
    \textbf{a}, Constant-height current image of two junctions with extra H atoms ($V=8$ mV). 
    \textbf{b}, Image with same conditions as in a after the removal of the extra H-atoms induced by tunneling electrons.
    The dehydrogenation processes were done over the ZZ$_1$ and PC$_2$ sites. 
    \textbf{c,d}, $dI/dV$ spectra taken over PC$_1$ and ZZ$_2$ regions (indicated in  \textbf{a} and  \textbf{b} respectively) before (black) and after (blue) the dehydrogenation processes.  Inset in  \textbf{c} shows the current during the process of dehydrogenation.}
\end{center}
\end{figure}

The presence of extra H atoms in Type 1 and 2 junctions \li{was confirmed by  electron induced H-atom removal experiments. Figure 4 \textbf{a} shows a structure formed by three chGNRs connected via Type 1 and 2 junctions.} Accordingly, their $dI/dV$ spectra (black curves in Figs.~4\textbf{c,d}) show a Kondo resonance at the PC$_1$ and ZZ$_2$ regions.  We placed the STM tip on top of the opposite  sites ZZ$_1$ and PC$_2$, and raised the positive sample bias well above 1 V. A 
step-wise decrease of the tunneling current indicated the  removal of the extra H atom (inset in Fig.~4\textbf{c}). The resulting junction appeared with double bright regions in 
low-bias images (Fig.~4\textbf{b}), and the PC$_1$ and ZZ$_2$ spectra turned into  dI/dV steps characteristic of Type 3  junctions (blue curves in Figs.~4\textbf{c,d}). Thus, the removal of H atoms  activated the magnetic moment of the initially unpolarized ZZ$_1$ and PC$_2$ sites, converting Type 1 and 2 junctions into Type 3, \np{and switching the total spin of the junction from spin   \textonehalf\ to zero.}

\begin{figure}
\begin{center}
	\centering
	\includegraphics[width=\columnwidth]{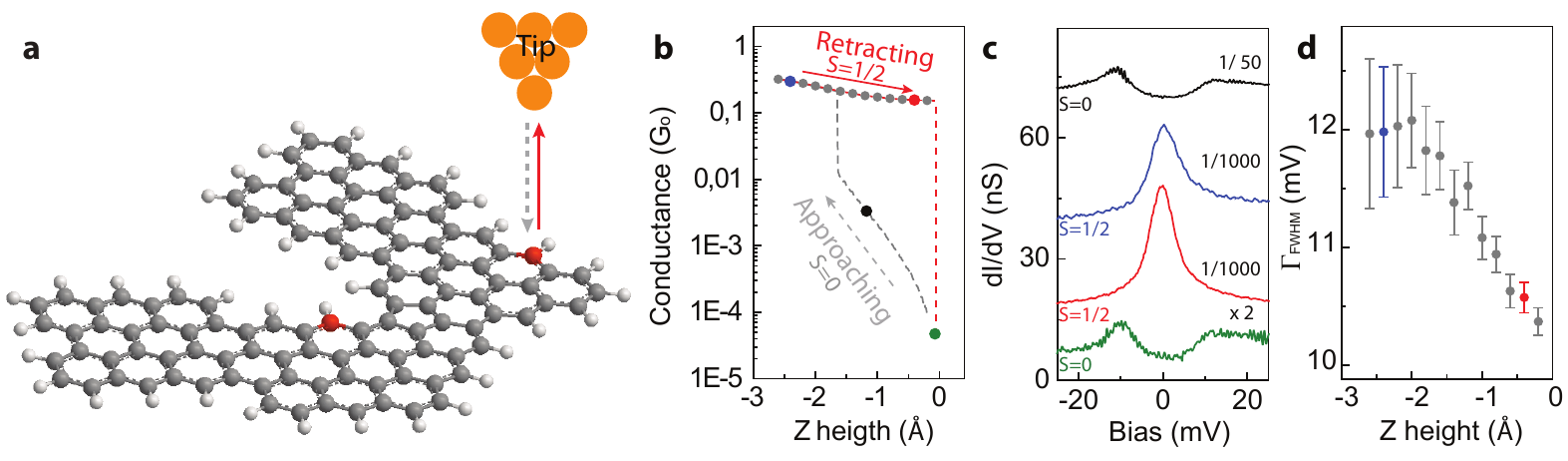}
	\caption{\textbf{Kondo effect from the spin embedded in a lifted chGNR junction}.
    \textbf{a}, Schematics of the process where the tip of the STM is first approached to the ZZ site of	a Type 3 junction (gray dashed arrow) and then retracted to lift the junction away from the substrate (red arrow), resulting in a suspended junction between tip and substrate.
    \textbf{b}, Simultaneously recorded conductance curve ($V=-50$ mV) during the approach, jump to contact and lift processes.
    \textbf{c}, $dI/dV$ spectra recorded at the specific heights indicated with  colored points on the curves in \textbf{b}. 
    \textbf{d}, Full widths at half maximum (FWHM) of spectra acquired in the retraction process (points in \textbf{c}), extracted from a fit using the Frota function\cite{Frota1992}. }
\end{center}
\end{figure}

\np{The magnetic state of the junction was also changed by creating a contact between the STM tip apex and a radical site. In  the experiments shown in Figure 5, the STM tip was approached to the  ZZ sites of a Type 3 junction. A  step in the conductance-distance plot (Fig.~5\textbf{b}) indicated the formation of a contact.  The created tip-chGNR contact could be stretched up to 3 \AA\ before breaking (retraction step  in Fig.~5\textbf{b}), signaling  that a chemical bond was formed. 
$dI/dV$ spectra recorded before the bond formation (black point in Fig.~5\textbf{c}) shows the split-peak feature of Type 3 junctions (black spectrum  in Fig.~5\textbf{c}). After the bond 
formation (blue and red points in Fig.~5\textbf{b}), the spectra changed to  show Kondo resonances  (blue and red spectra  in Fig.~5\textbf{c}),  persisting during contact retraction until the bond-breaking step, where double-peak features are recovered (green spectrum  in Fig.~5\textbf{c}). 
The formation of a tip-chGNR bond thus removed the spin of the ZZ site, and the transport 
spectra reflect the Kondo effect due to the remaining spin embedded in the junction. 
If the STM tip contacts instead the ZZ radical site of a Type 2 junction (shown in Supplementary Section 4) the initial Kondo resonance  disappears from the spectra, signaling the complete demagnetization of the junction.} The width of the Kondo resonance in the contacted  junctions  (blue and red plots in Fig.~5\textbf{c}) is significantly larger than in Type 1 and 2 cases, probably because it incorporates scattering with tip states\cite{Jasper-Tonnies2017,Choi2017}, and monotonously narrows as the contact is pulled apart (Fig.~5\textbf{d}). The survival of the Kondo effect in the contacted Type 3 junctions is a remarkable outcome of our experiments, which demonstrate the addressability of such localized magnetic moments in graphene nanostructure devices.

\section*{Reference}
\bibliographystyle{biblatex-nature}
\bibliography{thebibliography}

\begin{methods}

	\subsection{Sample preparation and experimental details.}
	The experiments were performed on two different scanning tunneling microscopes
	(STM) operating in ultra-high vacuum. A commercial JT STM (from specs) operated
	at 1.2 K with a magnetic field up to 3 Tesla was used to measure the temperature
	and magnetic field dependence of the Kondo resonance, while other experiments
	were done with a home made STM operating at 5 K. Both setups allow in situ
	sample preparation and transfer into the STM. The Au(111)
	substrate was cleaned in UHV by repeated cycles of Ne$^{+}$ ion sputtering and
	subsequent annealing to 730 K. The molecular
	precursor (2,2'-dibromo-9,9'-bianthracene) was sublimated at 170 $^{\circ}$C
	from a Knudsen cell onto the clean Au(111) substrate kept at room temperature.
	Then the sample was first annealed at 200 $^{\circ}$C for 15 minutes in order to
	induce the polymerization of the molecular precursors by Ullmann coupling, then
	the sample was annealed at 250 $^{\circ}$C for 5 minutes to trigger the
	cyclodehydrogenation to form chiral graphene nanoribbons (chGNRs). A last step
	annealing at 350 $^{\circ}$C for 1 minute created nanostructure junctions.
	A tungsten tip functionalized with a CO molecule was used for high resolution
	images. All the images in the manuscript were acquired in constant height mode, at
	very small voltages, and junction resistances of typically 20 M$\Omega$. The $dI/dV$
	signal was recorded using a lock-in amplifier with a bias modulation of
	$V_\mathrm{rms}=0.4$ mV at 760 Hz.\\
	
	\subsection{Simulations.}
	
	We performed calculations with the SIESTA implementation\cite{SIESTA} of density functional theory (DFT). Exchange and correlation (XC) were included within either 
	the local (spin) density approximation (LDA)\cite{LDA} or the generalized gradient approximation (GGA)\cite{PBE}. We used a 400 Ry cutoff for the real-space grid integrations and a double-zeta plus polarization (DZP) basis set
	generated with an 0.02 Ry energy shift for the cutoff radii. The molecules, represented with periodic unit cells, were separated by a vacuum of at least 10 {\AA} in any direction. The electronic density was converged to a stringent criterion of $10^5$.
	The force tolerance was set to 0.002 eV/{\AA}. Here is a description of a specific method used. To complement the DFT simulations described above we also performed 
	simulations based on the mean-field Hubbard (MFH) model, known to provide 
	a good description for carbon $\pi$-electron systems\cite{Fernandez-Rossier2007,Yazyev2010,Yazyev2011,Carvalho2014,Hancock2010}.

\end{methods}


\begin{addendum}
 \item We thank Manuel Vilas-Varela for the synthesis of the chGNR molecular precursor. We are indebted to Carmen Rubio, Dimas G. de Oteyza, Nestor Merino, Nicol\'as Lorente, Aran Garc\'{\i}a Lekue, and Daniel S\'anchez Portal for fruitful discussions. \\
We acknowledge financial support from Spanish AEI 
(MAT2016-78293-C6, FIS2017-83780-P, and the Maria de Maeztu Units of Excellence Programme 
MDM-2016-0618),  the Basque Government (Dep.~Industry, Grant PI-2015-1-42), the EU project 
PAMS (610446), the Xunta de Galicia (Centro singular de investigaci\'on de Galicia 
accreditation 2016-2019, ED431G/09),   and the European Regional Development Fund (ERDF).\\ 
 \item[Author Contributions] J.L,  and J.I.P. devised the experiment.  D.P. designed the  
organic synthesis of the chGNR molecular precursor. J.L. realized the measurements. S.S. 
and T.F. did the theoretical simulations. All the authors discussed the results.
J.L., T.F., and  J.I.P. wrote the  manuscript. \\
 \item[Competing Interests] The authors declare that they have no
competing financial interests.

\end{addendum}

\end{document}


\date{\today}
 \maketitle

\begin{affiliations}
	\item   CIC nanoGUNE, 20018 Donostia-San Sebasti\'an, Spain 
	\item \ Donostia International Physics Center (DIPC), 20018 Donostia-San Sebasti\'an, Spain
	\item Centro de F{\'{\i}}sica de Materiales 	MPC (CSIC-UPV/EHU),  20018 Donostia-San Sebasti\'an, Spain
	\item Centro Singular de Investigaci\'on en Qu\'imica Biol\'oxica e Materiais Moleculares (CiQUS), and Departamento de Qu\'imica Org\'anica, Universidade de Santiago de Compostela, Spain
	\item Ikerbasque, Basque Foundation for Science, 48013 Bilbao, Spain 
	\date{\today}
\end{affiliations}

 \setstretch{0.0} 
\baselineskip0pt
\tableofcontents

\newpage
\baselineskip18pt

\section{Reaction pathway for the formation of the GNR junctions}

\begin{figure}[b!]
\begin{center}
\includegraphics[width=0.99\textwidth]{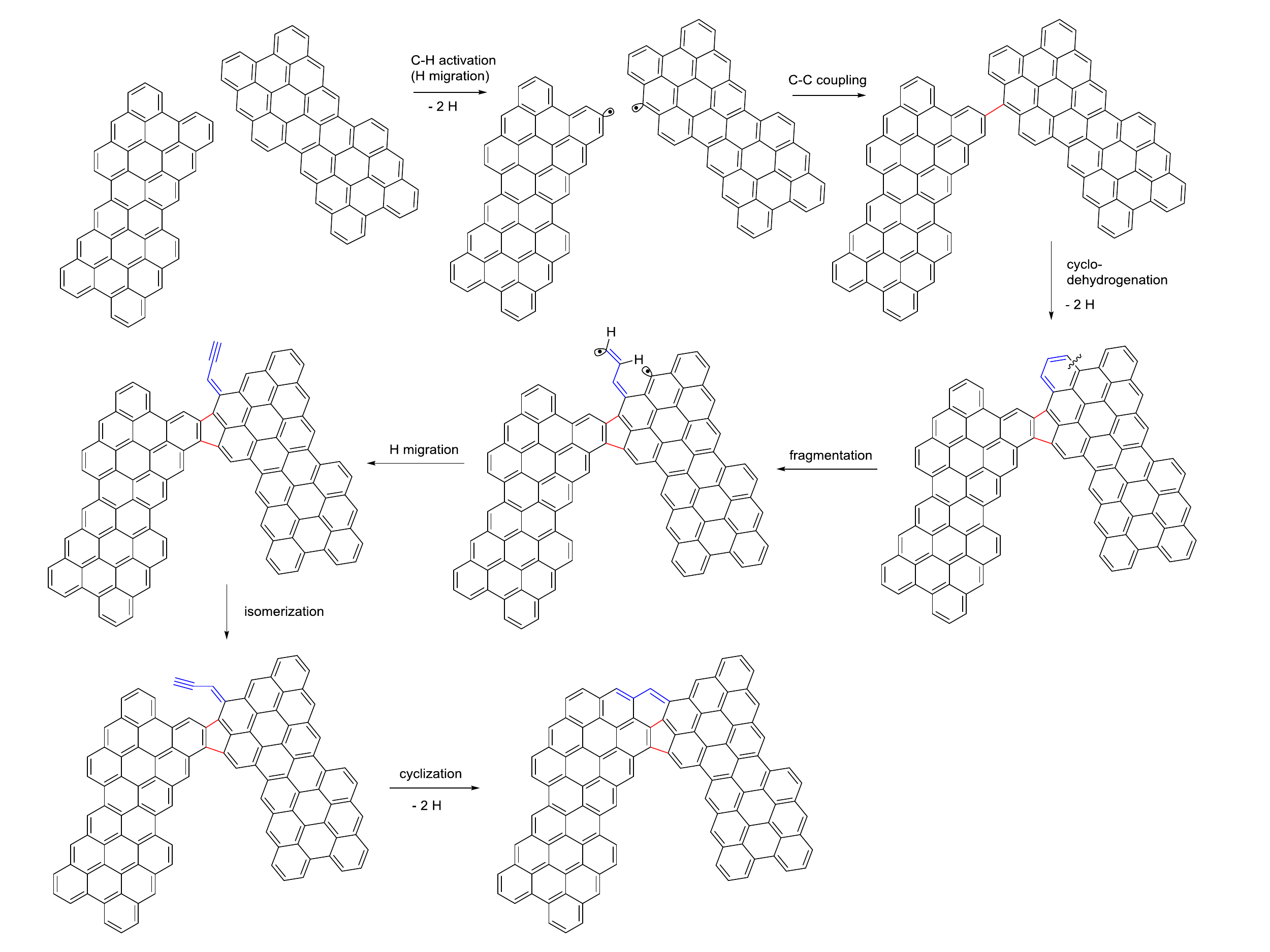}
\bfcaption{Proposed reaction pathway for the formation of the graphene nanostructures. }
{}
 \label{fig:pathway}
 \end{center}
 \end{figure}

Figure \ref{fig:pathway} shows a plausible mechanism for the formation of the GNR junctions. 
First, CH activation would form the corresponding sigma-radicals which could evolve by C-C coupling followed by cyclodehydrogenation to obtain a five-membered ring (in red). 
Then, fragmentation of a benzene ring, followed by H migration could lead to a terminal alkyne (in blue).
Finally, isomerization of a double bond, followed by cyclization of the terminal alkyne with a bay region of the GNR, could afford the final GNR junction.

\clearpage
\section{Statistics of different types of junctions}

\begin{figure}[b!]
\begin{center}
\includegraphics[width=0.8\textwidth]{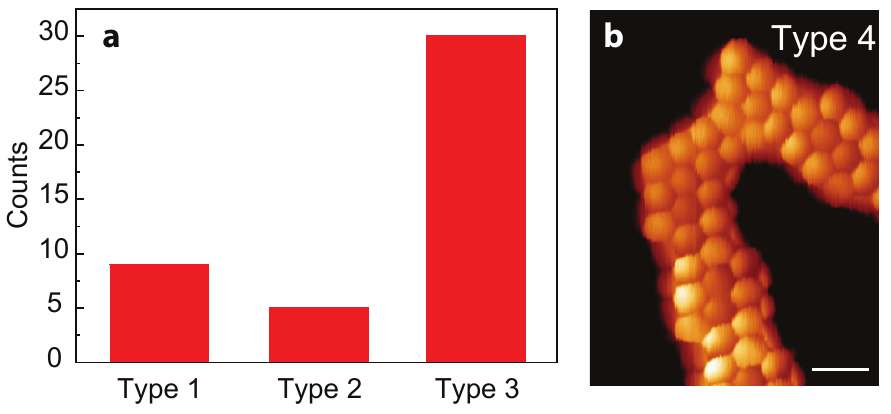}
\end{center}
\caption{\textbf{Statistics of different types of junctions.} a, Bar plot of the number of three types of junctions studied in the main text. b, Constant-height current image ($V=2$ mV, scale bar 0.5 nm) shows another type of junctions with both radicals are passivated by H atoms, which is classified as Type 4 junction.
\label{fig:statistics}
}
\end{figure}

In the main text we studied three different nanostructure junctions (Figs.~1 and 2). 
Here we show the frequency statistics of the three types. From
the 45 different junctions studied, 30 of them are identified as
Type 3 junctions, while 9 and 5 of them are identified as Type 1 and Type 2
junctions respectively (\Figref{fig:statistics}a). From this statistics, we deduced that the ZZ sites are more favorable to incorporate an extra hydrogen atom and get passivated (22\% of the radicals). The PC sites had an extra atom only in 13\% of the cases. The overall percentage of H-passivation observed here is comparable to the  value found at the termination of armchair GNRs\cite{Talirz2013}, 
which was happened in a 15\%  of the occasions. 

In one of the 45 nanostructures inspected, we  observed a junction with same backbone structure as the other three types, but with neither bright sites, nor zero-bias features in the spectra. \Figref{fig:statistics}b shows a  constant height current image  of the junction, where its ring structure can be now nicely resolved. It corresponds to  the carbon backbone sketch in Fig. 1d of the main text but with two extra  H atoms saturating the radical sites.

\newpage


\section{Understanding the appearance of spins in the junctions from Clar's theory}

\begin{figure} [b!]
\begin{center}
\includegraphics[width=0.9\textwidth]{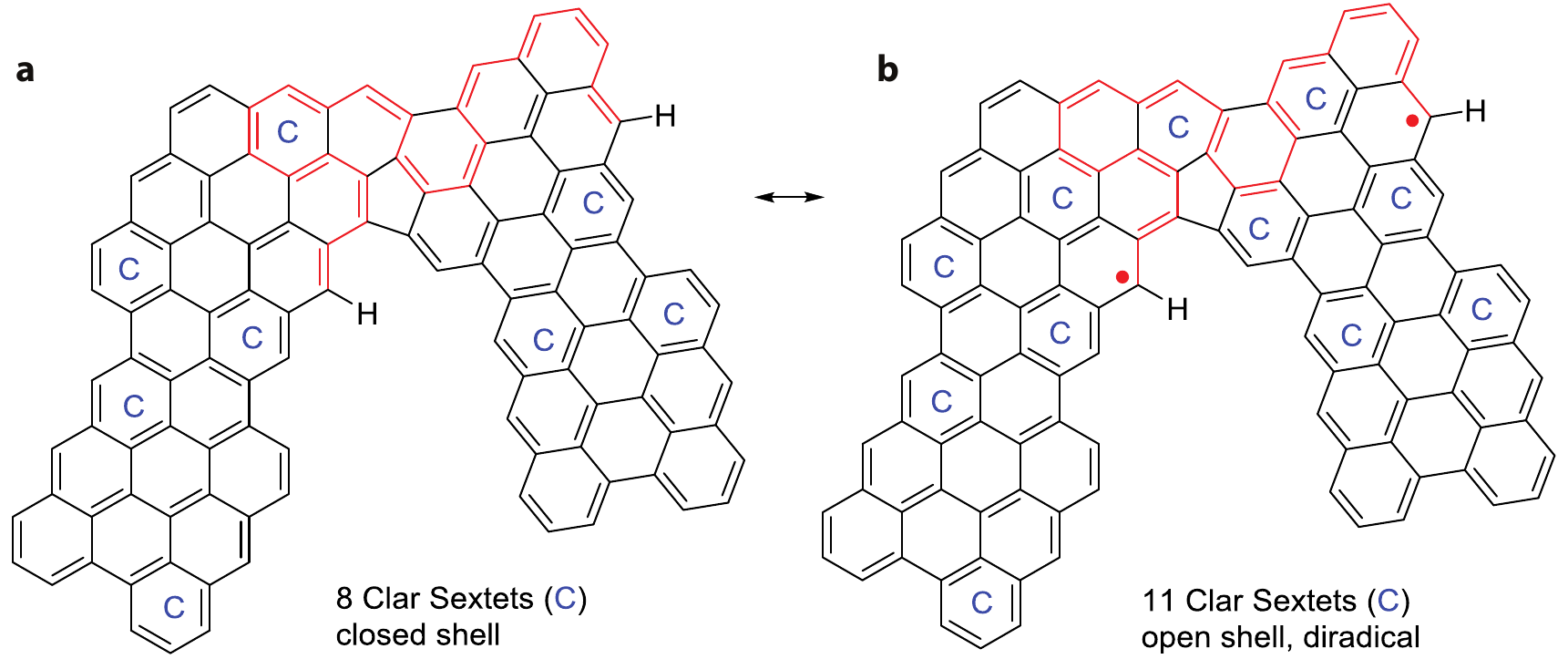}
\caption{\textbf{Understanding the appearance of spins in the nanostructures from Clar's theory. }
a, Close-shell resonant structure of a Type 3 nanostructure, with its  Clar's sextets indicated with "c".  
 b, Di-radical form of the model structure in a, hosting three additional sextets.  Only the H atom of the sites becoming radicals is shown in this model. Structure b is the energetically most stable if  the energy gain by incorporating the additional sextets  compensates the energy cost for creating the pair of radicals. In this case, the di-radical form is expected to show spin-polarization at the radical sites. 
  }
\label{fig:Clar}
\end{center}
\end{figure}

In the main text, we used both DFT and Hubbard Mean Field simulations to show that the 
spins in Type 3 junctions comes from two in-gap states, simply occupied as a result of electron-election correlations.
An alternative and intuitive chemical picture behind the emergence of singly occupied radical states  can be drawn bearing in mind Clar's aromatic pi-sextet rule \cite{Sola2013}. 
Figure \ref{fig:Clar} shows two possible resonance structures for the GNR junction: the closed shell structure a, with 8 Clar sextets, and the open shell structure b, with 11 Clar sextets and two radicals  at the PC and ZZ sites.
The dominance of resonance structure b in our experiments means that the energy required to create the two unpaired electrons (radicals)  in structure b is compensated by the stabilization provided by the presence of three additional Clar sextets.
In fact, the radical sites can delocalize towards the two second neighbor edge carbon atoms, agreeing with the carbon sites with high density of states shown in Fig.~1e in the main text.
Hence, this phenomenological model can qualitatively explain the spontaneous appearance of spin in the nanostructures.

\newpage


\section{Transport spectra of Type 2 junctions}

\begin{figure}[b!]
\includegraphics[width=\textwidth]{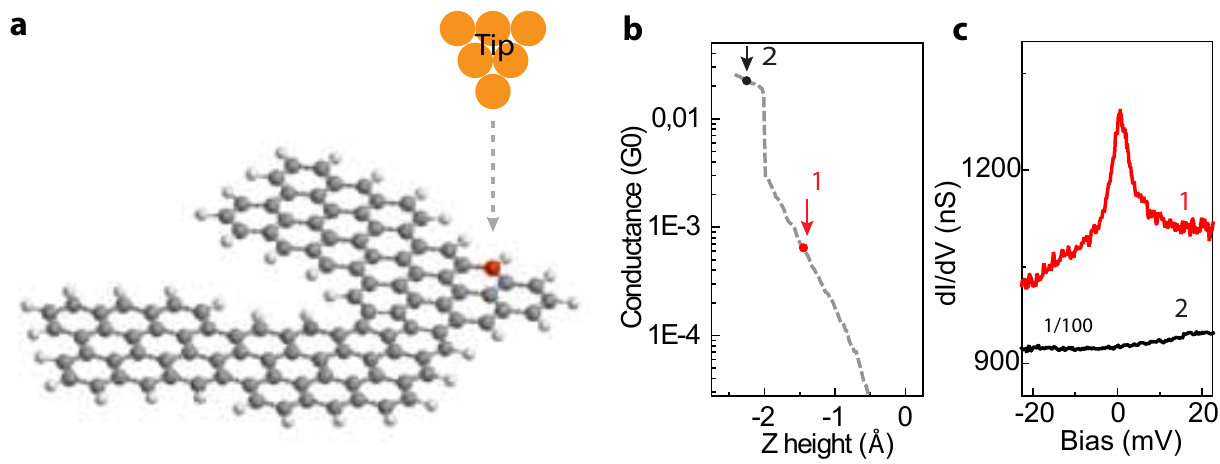}
\bfcaption{Transport properties of the lifted junctions without radical.}
{a, Schema illustrating the process, when the STM tip was approached to the
ZZ radical of a Type 2  junction (gray dashed arrow) to form a contact. 
b, Simultaneously recorded conductance curve ($V=-50$ mV) during the process in a. Red and
black dots indicate the vertical positions at whcih 
$dI/dV$ spectra in c were taken.  
 }
\label{fig:transport}
 \end{figure}

In Fig. 5 of the main text,  we studied the behaviour of a   spin localized at the PC site in a transport measurement,  when a Type 3  junction was contacted with the STM tip at the neighbour ZZ site and lifted. Electrons injected through the conjugated backbone reproduced the Kondo resonance observed in tunneling regime. 
For comparison, here we show similar transport measurement for a Type 2 junction, i.e. when there is no spin in the graphene nanostructure. As in the other case, the STM tip was approached to the radical at the ZZ site  to make a
bond between   nanostructure  and STM tip (illustrated in \Figref{fig:transport}). Before
bond formation, the characteristic Kondo resonance of Type 2 junctions is observed in the $dI/dV$ spectra (point 1 in the figure).   However, once the radical bonded to the tip  (signalled by the characteristic jump-to-contact step), the junction  bridged  tip and substrate and the Kondo resonance disappeared. This proves that the tip-radical contact quenches the magnetic moment of this site, as presumed in Fig. 5. It also proves that the Kondo feature observed in Fig. 5 for the lifted junction correspond to the PC spin embedded in the cGNR junction.   

\newpage


\section{DFT simulations}
We performed calculations with the SIESTA implementation
\cite{SIESTA} of density functional theory (DFT). 
Exchange and correlation (XC) were included within either 
the local (spin) density approximation (LDA) \cite{LDA} or
the generalized gradient approximation (GGA) \cite{PBE}. 
We used a 400 Ry cutoff for the real-space grid integrations 
and a double-zeta plus polarization (DZP) basis set
generated with an 0.02 Ry energy shift for the cutoff radii.
The molecules, represented with periodic unit cells,
were separated by a vacuum of at least 10 {\AA} in any direction.
The electronic density was converged to a stringent criterion of $10^5$.
The force tolerance was set to 0.002 eV/{\AA}.
In Fig.~3 in the main text we report the GGA results.

\subsection{Role of exchange-correlation functional}

\begin{figure}[h!]
\includegraphics[width=\textwidth]{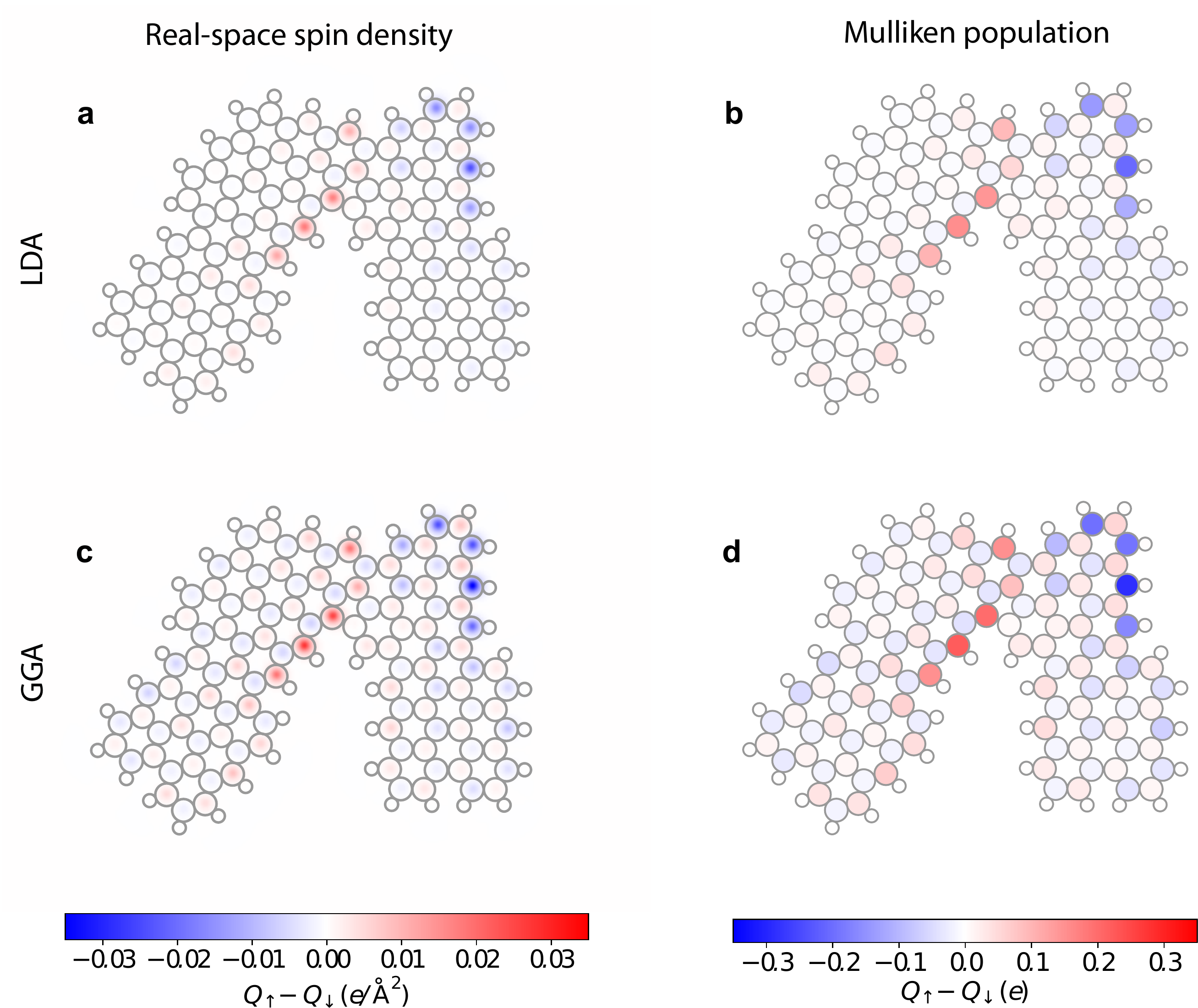}
\bfcaption{Spin polarization in the (2,2)-junction from DFT simulations.} 
{a, Real-space spin density calculated within LDA. 
b, Mulliken population analysis of the spin density calculated within LDA.
c,d Same as a,b but for GGA.
Panel d corresponds to Fig.~3a in the main text.
}
\label{fig:spin-pol-DFT}
\end{figure}

In \Figref{fig:spin-pol-DFT} we compare the calculated spin polarization for the 
generic (2,2) graphene nanojunction within both LDA \cite{LDA} and GGA \cite{PBE} XC approximations.
We also compare the real-space spin density with a Mulliken population analysis.
From \Figref{fig:spin-pol-DFT} it is clear that the emerging picture for the radicals
is robust among all four approaches. As expected,
the intensity of the spin polarization is more pronounced in GGA than in LDA.

\subsection{Energetically preferred hydrogen passivation sites}
\begin{figure}[t!]
\begin{center}
\vspace{-18mm}
\includegraphics[width=.65\textwidth]{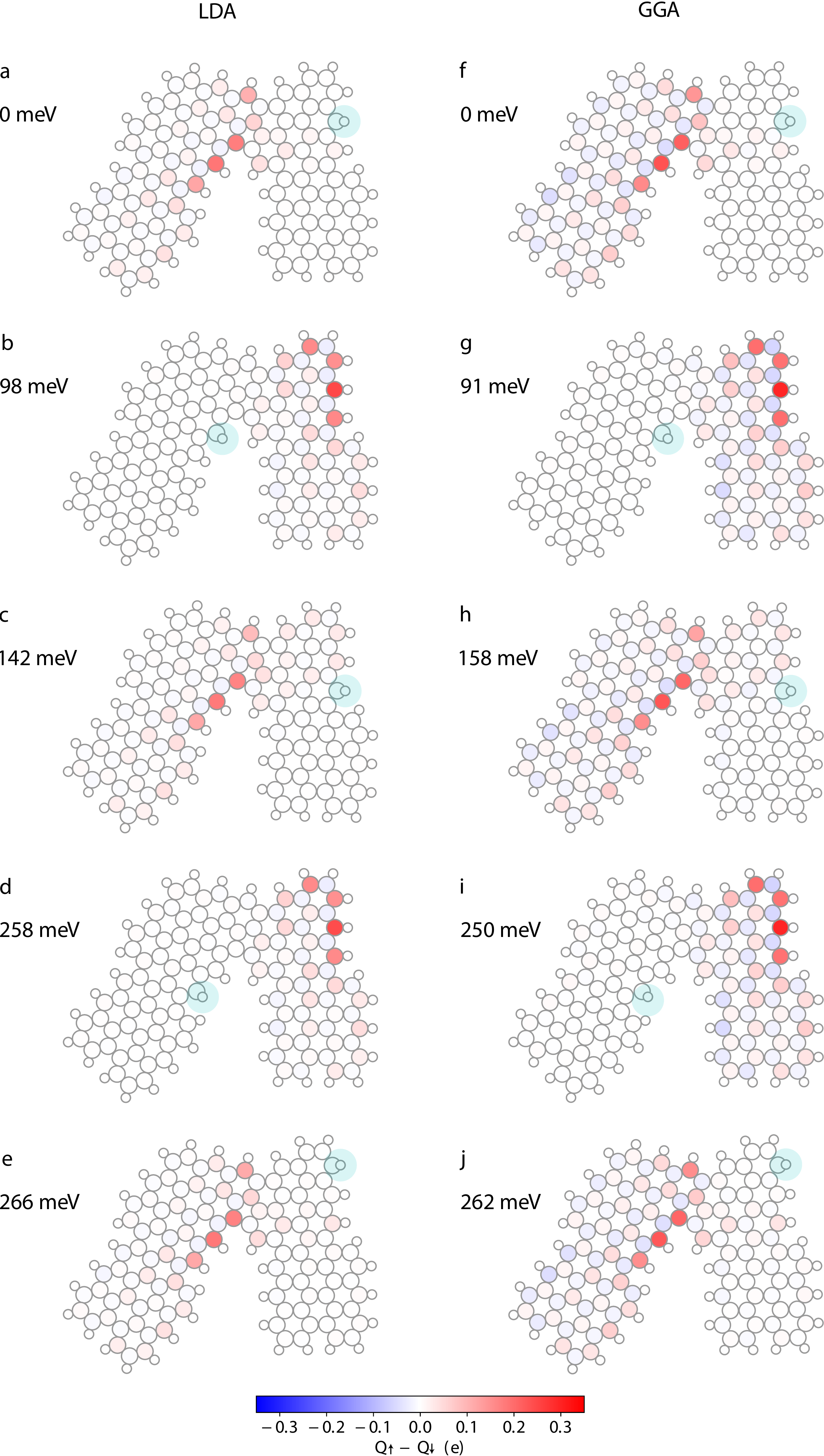}
\end{center}
\bfcaption{Energetics and spin polarization for five energetically preferred hydrogen passivation sites from DFT simulations.} 
{a-e, Results from LDA calculations via a Mulliken population analysis. The hydrogen passivation (blue circles) shown in a on the ZZ site is the most stable configuration, while the other sites are less energetically favoured with the energy differences quoted in each panel.
f-j, Same molecules as in a-e but results from GGA calculations.
Panels f,g correspond to Fig.~3b in the main text.
}
\label{fig:H-pasiv}
\end{figure}

In the main text we showed the hydrogen passivation on ZZ sites (Type 1 junctions)
and PC sites (Type 2 junctions).
In \Figref{fig:statistics} we report a higher probability of hydrogen passivation on ZZ sites than PC sites.
To quantitatively study this phenomenon, we analysed the energetics of different hydrogen passivations of the edges from DFT simulations.
The results are summarized in \Figref{fig:H-pasiv}.
We find that hydrogen passivation on the ZZ and PC sites are indeed the two most stable configurations, with the former being the energetically most favoured one.
This is in agreement with the experimental observations (\Figref{fig:statistics}).  

\clearpage


\section{Mean-field Hubbard model}
To complement the DFT simulations described above we also performed 
simulations based on the mean-field Hubbard (MFH) model, known to provide 
a good description for carbon $\pi$-electron systems \cite{Fernandez-Rossier2007,Yazyev2010,Yazyev2011,Carvalho2014,Hancock2010}. 
We describe the graphene nanostuctures with the following Hamiltonian for the $sp^2$ carbon atoms:
\begin{eqnarray}
H &=& -t_1 \sum_{\langle i,j\rangle,\sigma} (c_{i\sigma}^\dagger c_{j\sigma}+\mathrm{h.c.})
-t_2 \sum_{\langle\!\langle i,j\rangle\!\rangle,\sigma} (c_{i\sigma}^\dagger c_{j\sigma}+\mathrm{h.c.}) -t_3 \sum_{\langle\!\langle\!\langle  i,j\rangle\!\rangle\!\rangle,\sigma} (c_{i\sigma}^\dagger c_{j\sigma}+\mathrm{h.c.})\nonumber\\
&&+ U \sum_{\langle i}\big(n_{i\uparrow}\langle n_{i\downarrow}\rangle
+\langle n_{i\uparrow}\rangle n_{i\downarrow}
-\langle n_{i\uparrow}\rangle \langle n_{i\downarrow}\rangle \big)
\label{eq:MFH}
\end{eqnarray}
where $c_{i\sigma}$ ($c_{i\sigma}^\dagger$) annihilates (creates) an electron with spin $\sigma$ in the $p_z$ orbital centred at site $i$.
The first three terms describe a tight-binding model with 
hopping amplitudes $t_1$, $t_2$, and $t_3$ for the first, second, and third-nearest neighbour matrix elements
(defined in terms of interatomic distances  $d_1<1.6$\AA$<d_2<2.6$\AA$<d_3<3.1$\AA).
We follow the parameterizations of Ref.~\cite{Hancock2010} and consider
both a simple first-nearest neighbour (1NN) model with $t_1=2.7$ eV and $t_2=t_3=0$ as
well as a more accurate third-nearest neighbour (3NN) model
with $t_1=2.7$ eV, $t_2=0.2$ eV, and $t_3=0.18$ eV.

The term proportional to the empirical parameter $U$ accounts for the on-site Coulomb repulsion.
By comparison with first-principles simulations it has been established that DFT-GGA (DFT-LDA) are generally best reproduced when $U/t \approx 1.3$ (0.9) \cite{Yazyev2011}. Consistent with this, we find a good overall agreement with our experimental observations using $U\sim 3.5$ eV as analysed below.

The expectation value of the spin-resolved density operator $n_{i\sigma}=c_{i\sigma}^\dagger c_{i\sigma}$ is computed from the eigenvectors of $H$.
From the self-consistent solution of the Hamiltonian in Eq.~(\ref{eq:MFH}) we obtain the local spin density
from the charge difference $Q_{i\uparrow}-Q_{i\downarrow}$, with $Q_{i\sigma}=e \langle n_{i\sigma}\rangle$.
In units of $\mu_B$ the magnetization is $M_i=(n_{i\uparrow}-n_{i\downarrow})/2$.

We solve the mean-field Hubbard model using a custom-made Python implementation based on \textsc{sisl} \cite{SISL}.
In Fig.~3 in the main text we report the non-interacting single-particle wave functions in the 3NN model ($U=0$) as a basis to understand the open-shell electronic configurations obtained in DFT and MFH.


\newpage
\subsection{Band structure of infinite (3,1)cGNRs}

\begin{figure*}[b!]
\begin{center}
\includegraphics[width=.99\textwidth]{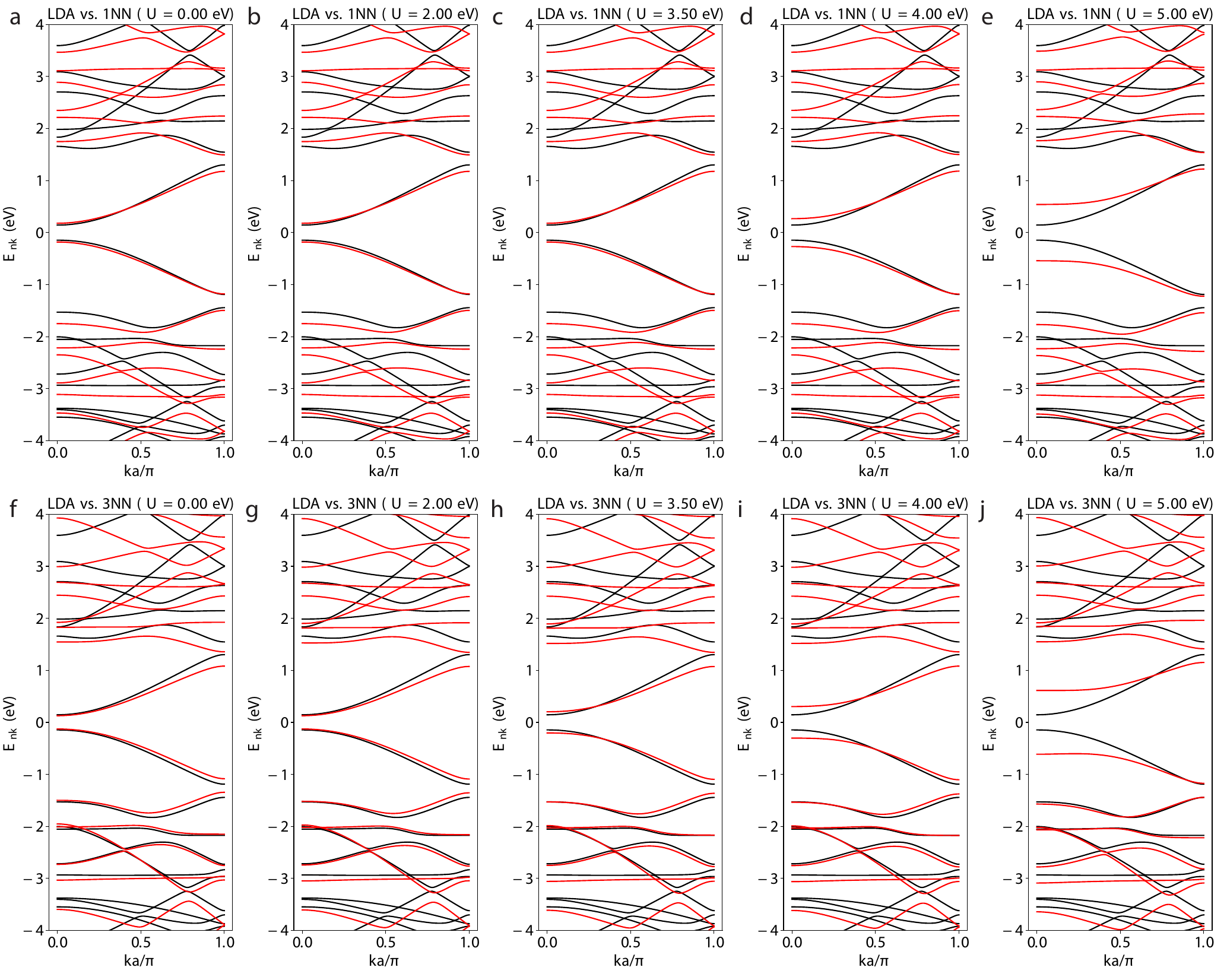}
\end{center}
	\bfcaption{Calculated 1D band structure for the (3,1)cGNR.}
	{We compare DFT-LDA (black lines) with the mean-field Hubbard model with different Coulomb repulsion $U$ (red lines) within
	either \pl{a-e} first-nearest neighbour couplings only ($t_1=2.70$ eV, top row)
	or \pl{f-j}  third-nearest neighbour couplings ($t_1=2.70$ eV, $t_2=0.20$ eV, and $t_3=0.18$ eV, bottom row).
    } 
\label{fig:bandsLDA}
\end{figure*}

\begin{figure*}[h!]
\begin{center}
 \includegraphics[width=.99\textwidth]{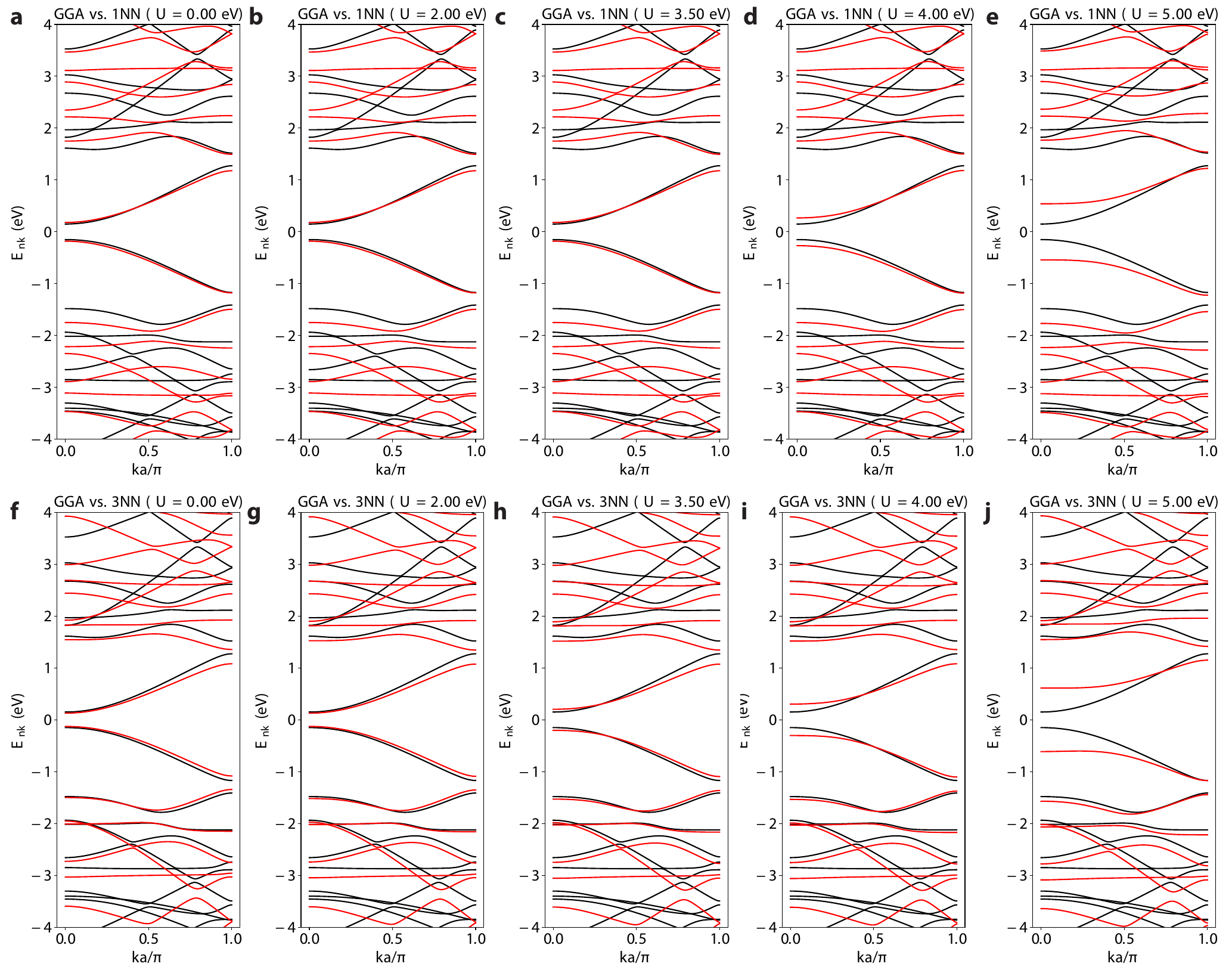}
\end{center}
	\bfcaption{Calculated 1D band structure for the (3,1)cGNR.}
	{We compare DFT-GGA (black lines) with the mean-field Hubbard model with different Coulomb repulsion $U$ (red lines) within 
	either \pl{a-e} first-nearest neighbour couplings only ($t_1=2.70$ eV, top row)
	or \pl{f-j}  third-nearest neighbour couplings ($t_1=2.70$ eV, $t_2=0.20$ eV, and $t_3=0.18$ eV, bottom row).
    } 
\label{fig:bandsGGA}
\end{figure*}

As shown in Figures.~\ref{fig:bandsLDA} and \ref{fig:bandsGGA}, both the first-neighbour and third-neighbour MFH models (red bands) provide a good description
for the 1D band structure of the (3,1) chiral graphene nanoribbon (cGNR) as compared to DFT calculations (black bands) obtained with \textsc{Siesta} \cite{SIESTA}.
Unlike DFT and the 3NN model, the simple 1NN model implies electron-hole symmetry of the bands. The low-energy part of the DFT band structure is generally very well reproduced with MFH using an on-site Coulomb repulsion of $U\leq 3.5$ eV.
Indeed this narrow cGNR is intrinsically non-magnetic, consistent with previous works \cite{Yazyev2011,Carvalho2014,Merino2017}.

\clearpage

\newpage
\subsection{Spin polarization}

\begin{figure*}[b!]
	\begin{center}
    \includegraphics[width=.6
    \textwidth]{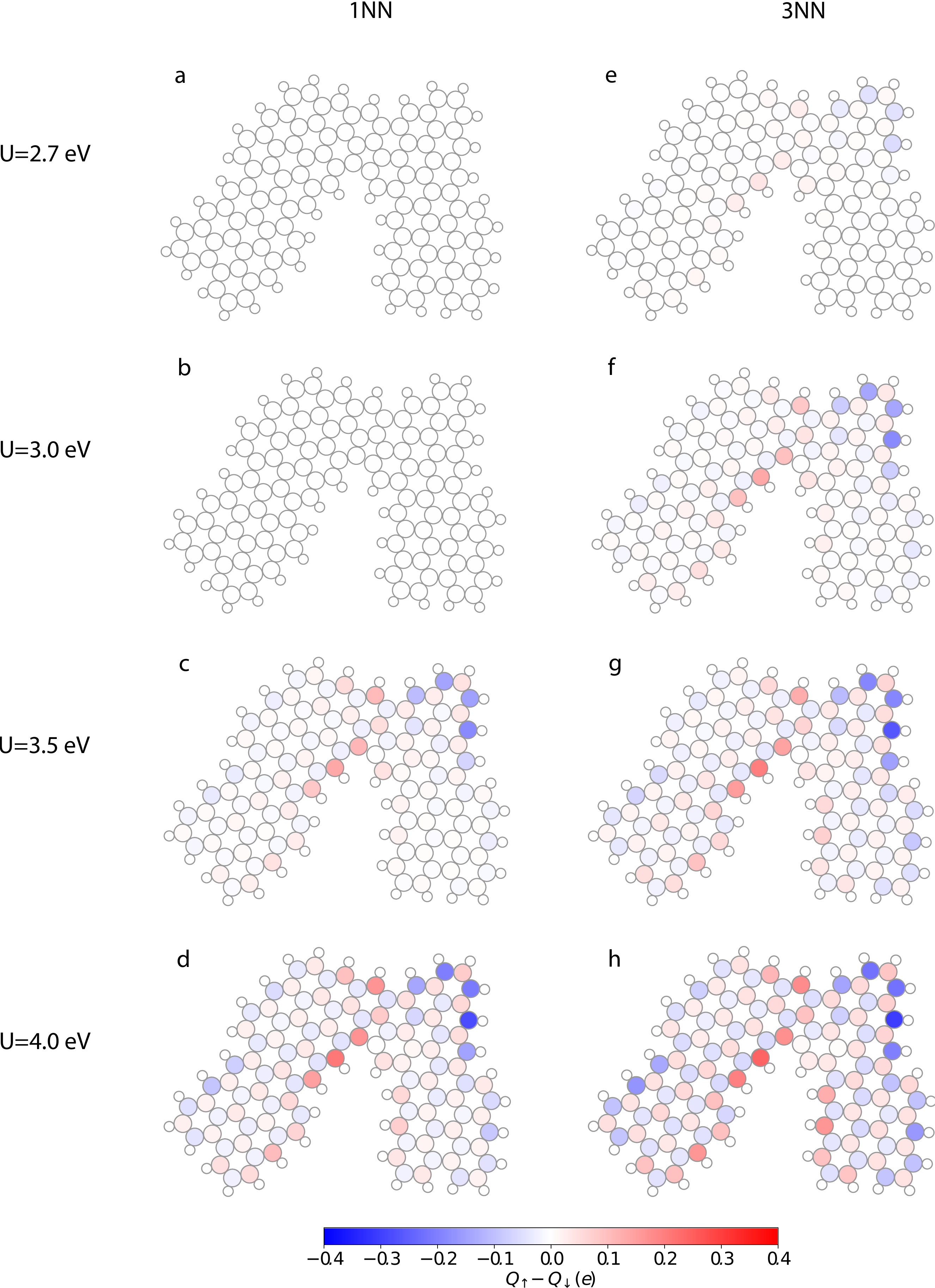}
	\end{center}
	\bfcaption{Spin density from MFH calculations.}
	{\pl{a-d} Spatial spin density, computed at site $i$ as $Q_{i\uparrow} - Q_{i\downarrow}$, for different values of the on-site Coulomb repulsion parameter $U$. \pl{e-h} Same as (a-d) but for the 3NN model.
    }
    \label{fig:pol-MFH}
\end{figure*}

In \Figref{fig:pol-MFH}  we examine the spatial spin density obtained with MFH for both 1NN and 3NN models and varying on-site Coulomb repulsion.
In the 1NN (3NN) model the onset of spin polarization occurs around $U=3.1$ ($2.7$) eV.
Compared with the DFT results in \Figref{fig:spin-pol-DFT} (and the band structures of the previous section) we conclude that the 3NN model with Coulomb repulsion of the order $U=3.5$ eV yields a very satisfactory agreement with DFT.

\clearpage
\subsection{Single-particle wave functions}

\begin{figure*}[b!]
	\begin{center}
    \includegraphics[width=.75\textwidth,trim=30 0 48 0]{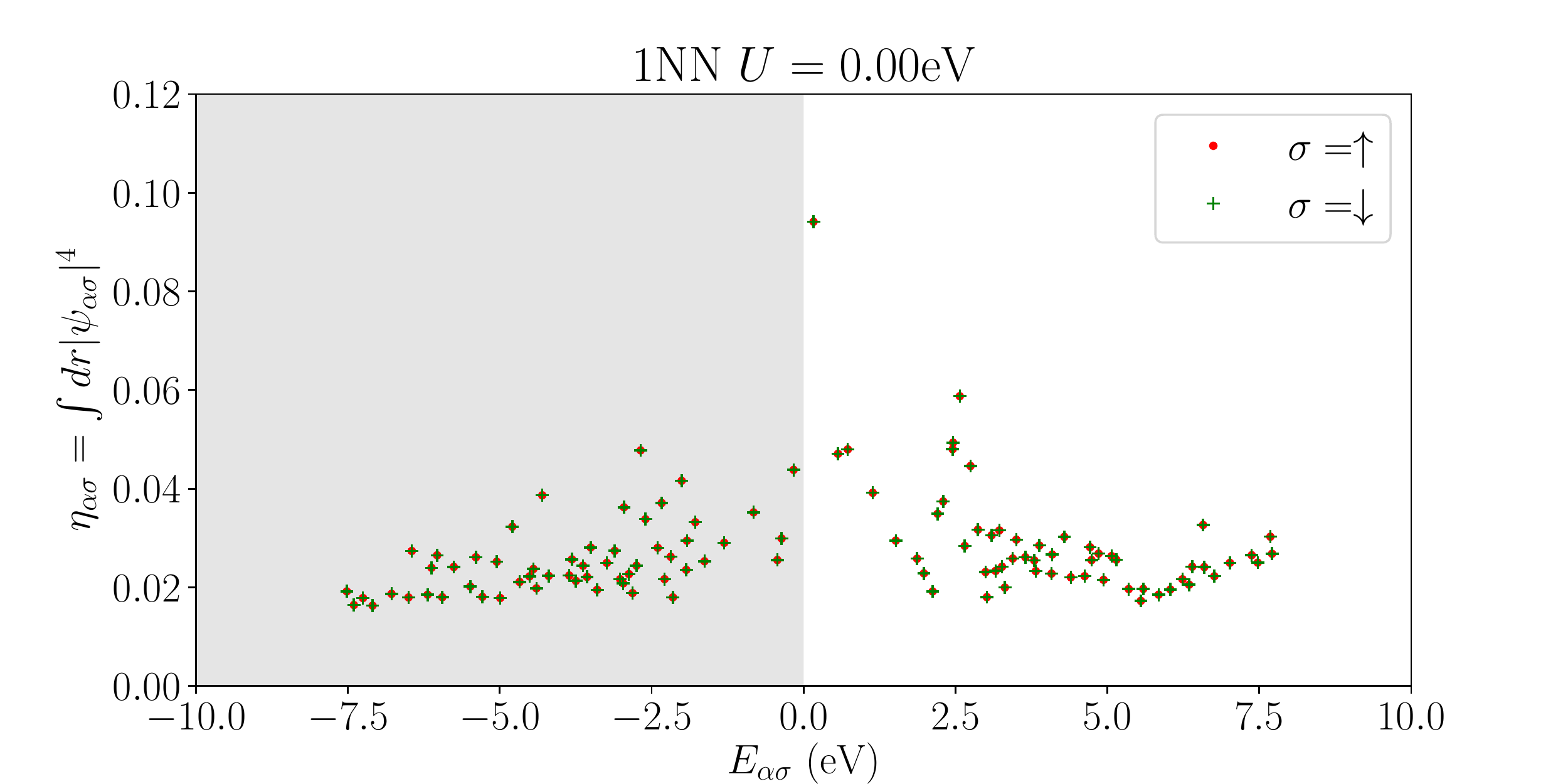}
    \includegraphics[width=.75\textwidth,trim=30 0 48 0]{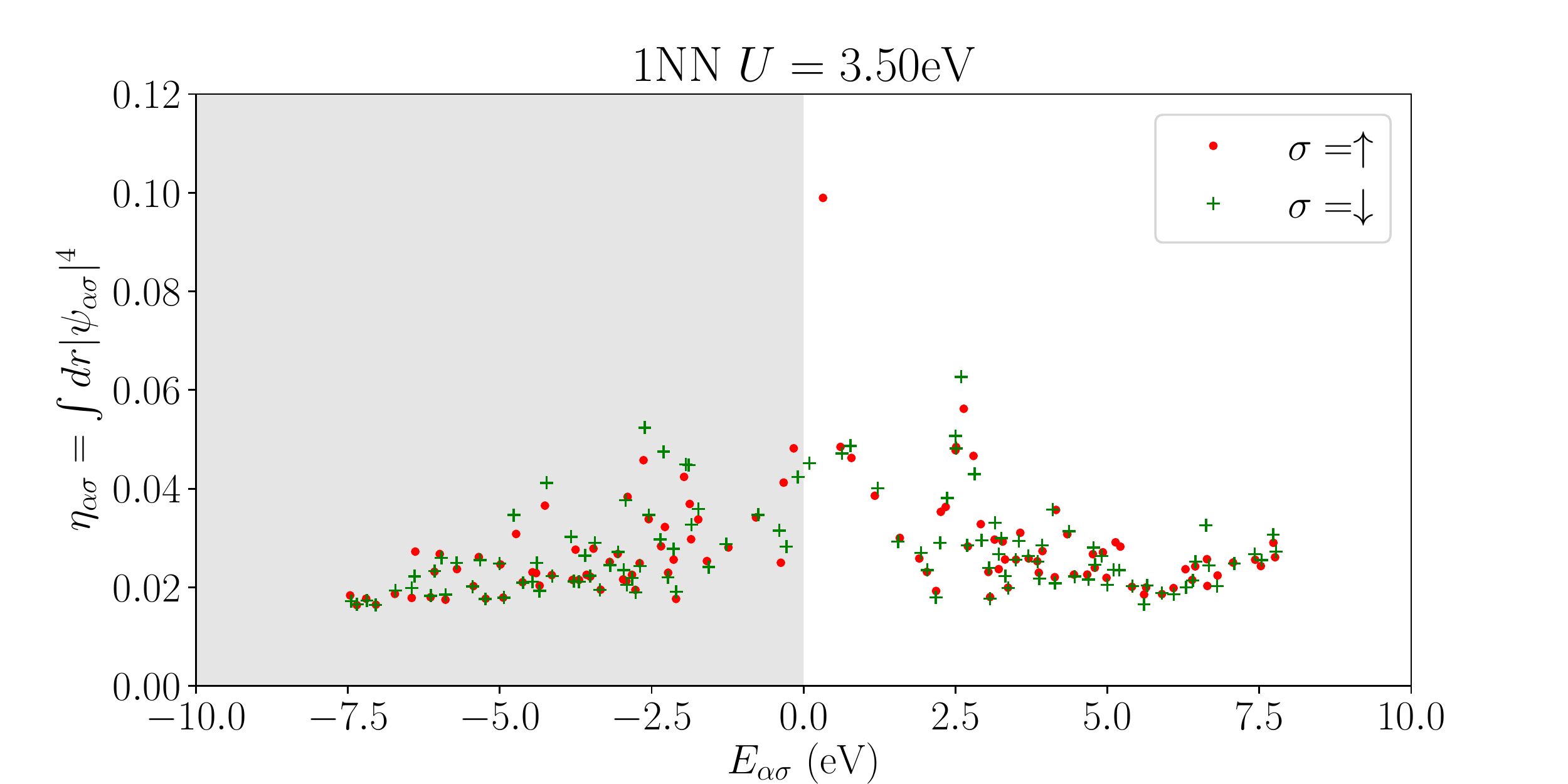}
    \end{center}
	\bfcaption{Single-particle orbital localization $\eta_{\alpha\sigma}$ versus single-particle energy $E_{\alpha\sigma}$ in the 1NN model.}
	{Here we consider two characteristic values of $U$ in both the 1NN and 3NN TB models.
	Among all states the LUMO orbital ($\sigma=\uparrow$ for finite $U$) is the most localized one.}
    \label{fig:1NNspectrum}
\end{figure*}

\begin{figure*}[b!]
	\begin{center}
    \includegraphics[width=.75\textwidth,trim=30 0 48 0]{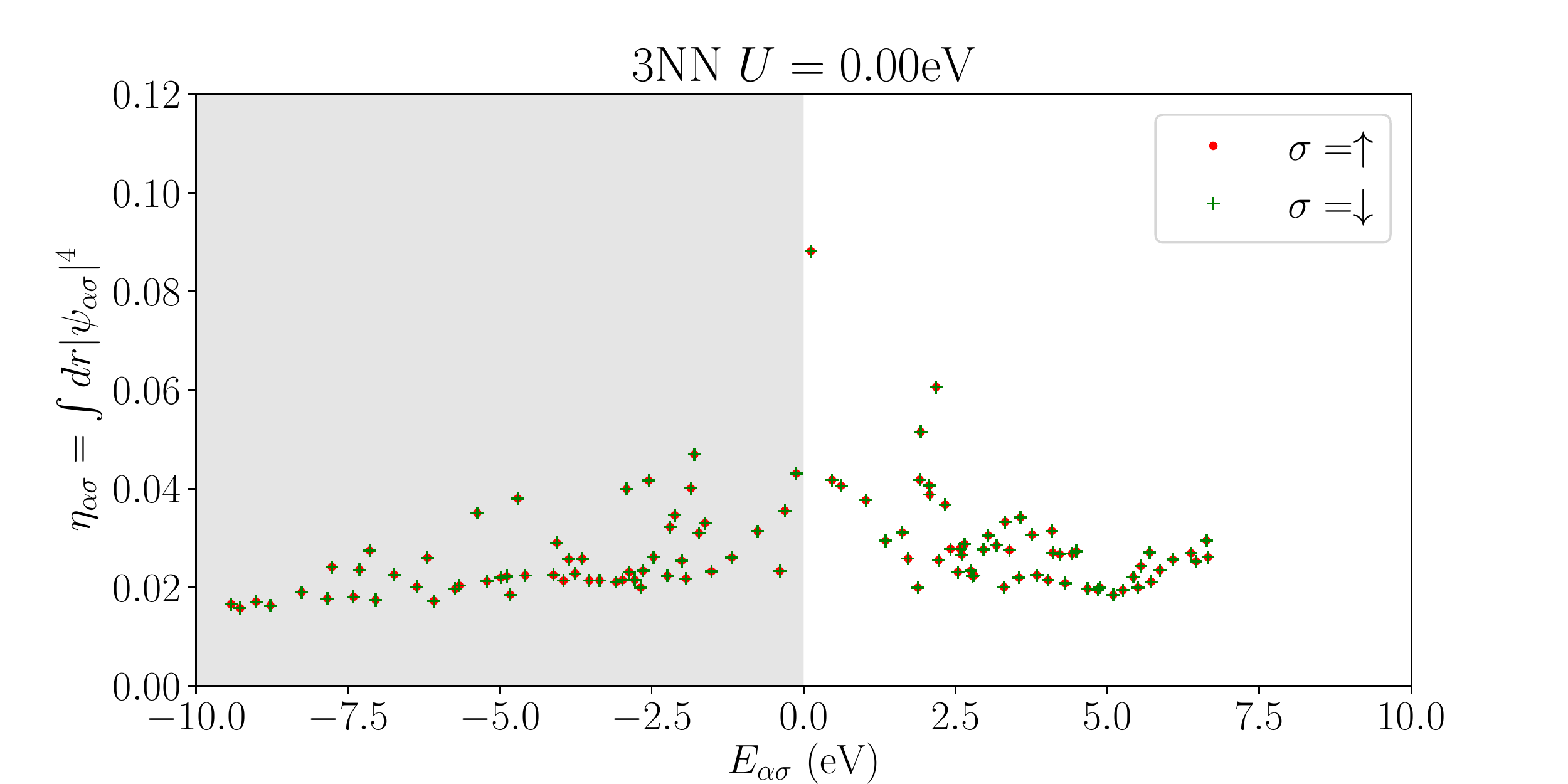}
    \includegraphics[width=.75\textwidth,trim=30 0 48 0]{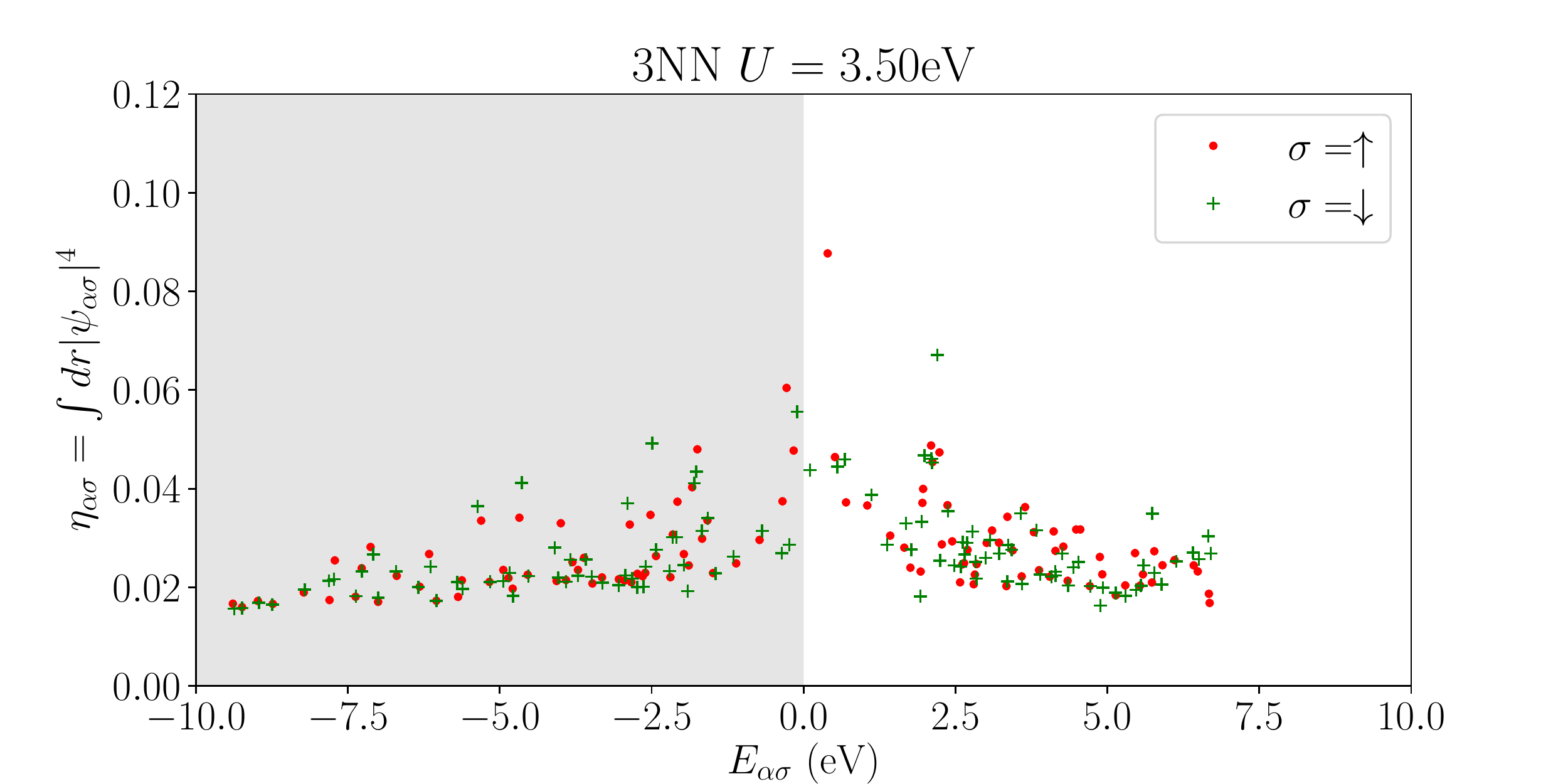}
    \end{center}
	\bfcaption{Single-particle orbital localization $\eta_{\alpha\sigma}$ versus single-particle energy $E_{\alpha\sigma}$ in the 3NN model.}
	{Here we consider two characteristic values of $U$ in both the 1NN and 3NN TB models.
	Among all states the LUMO orbital ($\sigma=\uparrow$ for finite $U$) is the most localized one.}
    \label{fig:3NNspectrum}
\end{figure*}

\begin{figure*}
	\begin{center}
	\includegraphics[width=1.\textwidth]{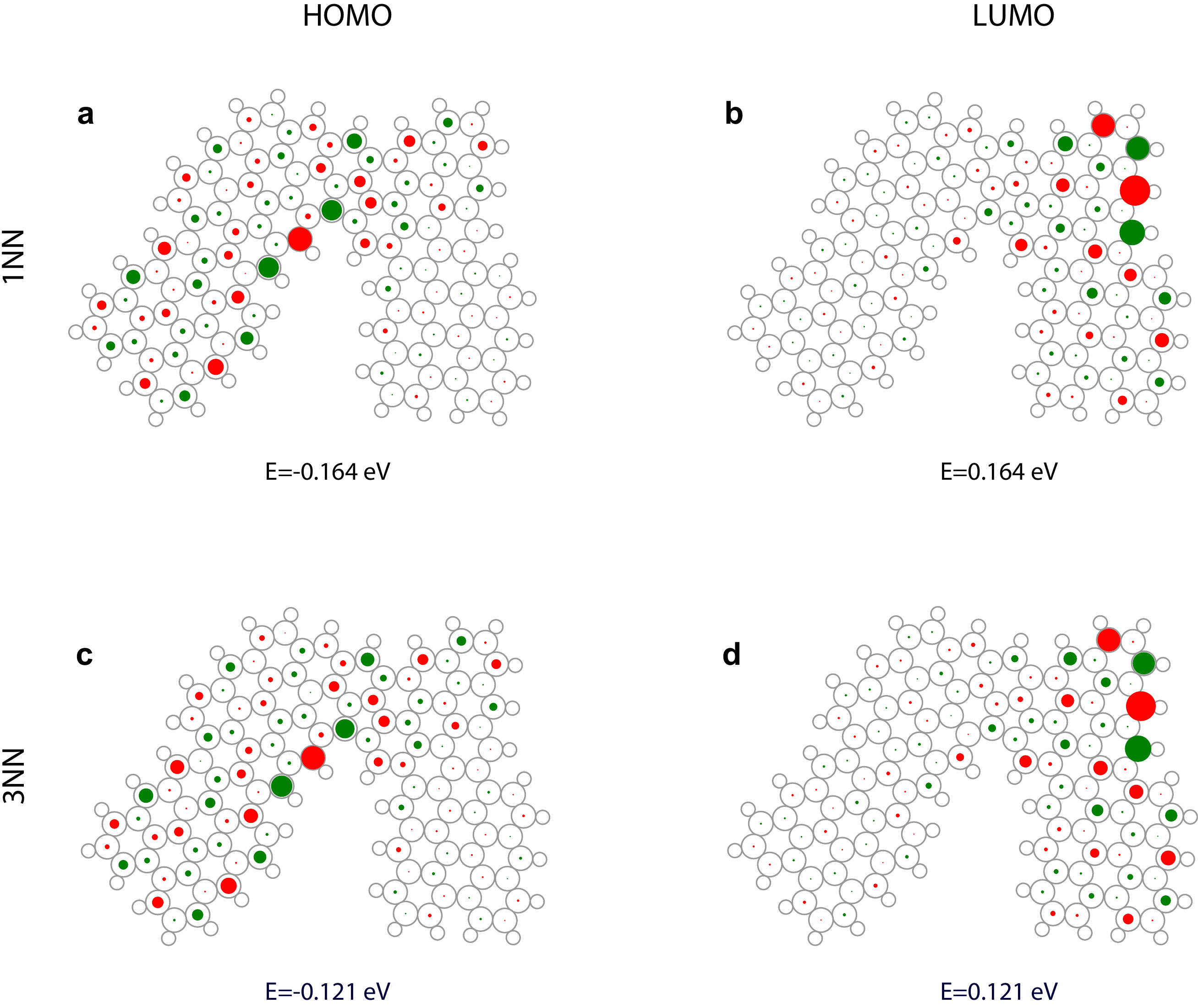}
    \end{center}
	\bfcaption{Single-particle wave functions from MFH with $U=0$ eV.}
    {a,b, Spatial distribution of the HOMO and LUMO wave functions 
     within the 1NN model.
    c,d, Same as a,b but within the 3NN model. 
    Panels c,d correspond to Fig.~3e in the main text.
    The single-particle energies relative to the midgap are stated below each plot.
    The size and color of the red-green circles reflect magnitude and phase of the wave function coefficients on each carbon atom, respectively.
    }
    \label{fig:WF0}
\end{figure*}

\begin{figure*}
    \begin{center}
    \includegraphics[width=1.\textwidth]{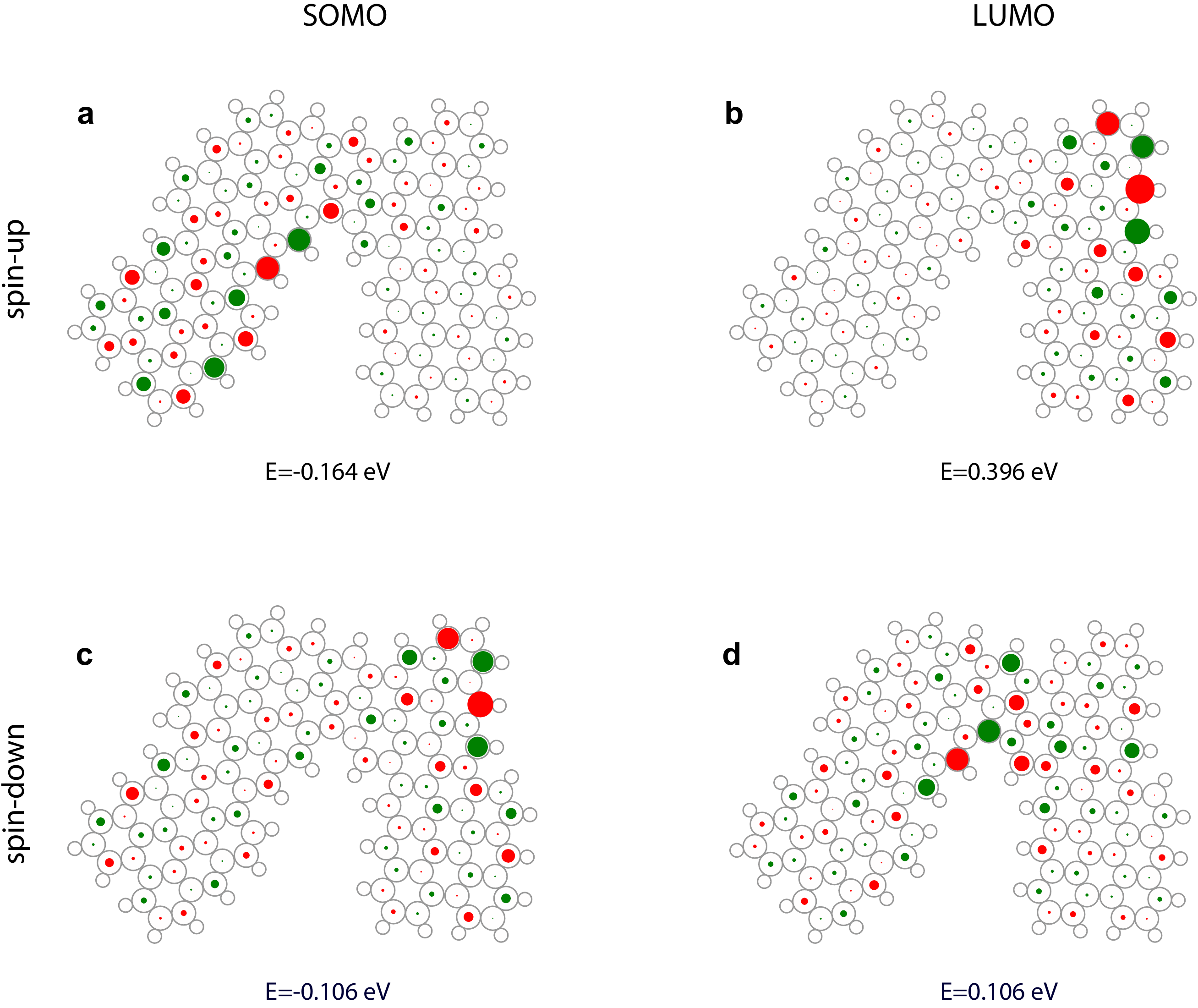}
	\end{center}
	\bfcaption{Single-particle wave functions from MFH within the 3NN model with $U=3.5$ eV.}
    {a,b, Spatial distribution of the SOMO and LUMO wave functions
    for the spin-up electrons, respectively.
    c,d, Same as a,b but for the spin-down electrons.
    The single-particle state energies relative to the midgap are stated below each plot.
    The size and color of the red-green circles reflect magnitude and phase of the wave function coefficients on each carbon atom, respectively.
     }
    \label{fig:WFU}
 \end{figure*}

In \Figref{fig:1NNspectrum} and \Figref{fig:3NNspectrum} we analyze the eigenspectrum of
energies and states in both the non-interacting ($U=0$) and interacting ($U=3.5$ eV) MFH Hamiltonians. 
The degree of spatial localization of each state is computed as $\eta_{\alpha\sigma}=\int dr |\psi_{\alpha\sigma}|^4$, also denoted the inverse participation ratio \cite{Ortiz2017}.  The HOMO/LUMO wave functions, shown in \Figref{fig:WF0} for $U=0$ 
and in \Figref{fig:WFU} for $U=3.5$ eV, are concentrated around the radical sites of the structure. 

\clearpage
\subsection{Singlet-triplet excitations}

\begin{figure}[b!]
	\begin{center}
		\includegraphics[width=.7\textwidth]{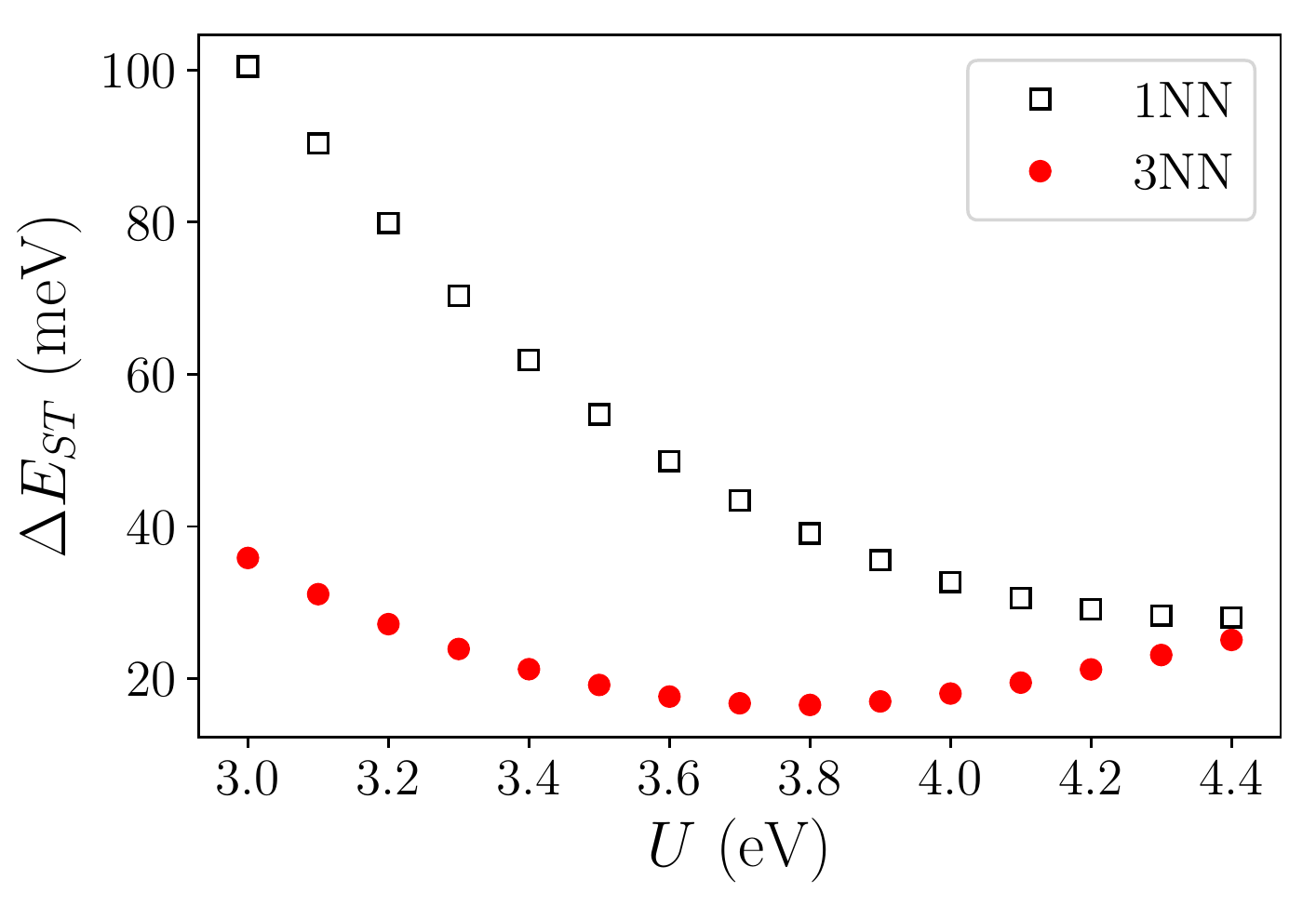}
	\end{center}
	\bfcaption{Singlet-triplet excitation energy $\Delta E_{ST}$ versus $U$.}
	{We consider here the prototype molecule $(L,R)=(2,2)$ within both the 1NN and 3NN models. The curves increase monotonically to the HOMO-LUMO gap $\Delta E_\mathrm{H-L}=0.328$ ($0.242$) eV in the limit $U\rightarrow 0$ for the 1NN (3NN) model.
	}
\label{fig:sing-trip}
\end{figure}

From \Figref{fig:pol-MFH} we have established that $U=3.5$ eV yields a good description for these nanostructures as compared with DFT.
As an approximation to the true singlet-triplet excitation energy $J$, we 
computed the mean-field energy difference $\Delta E_{ST}$ between
the converged electronic configurations with $n_{\uparrow}=n_{\downarrow}$ and $n_{\uparrow}=n_{\downarrow}+2$.
In \Figref{fig:sing-trip} we explore the variation of
$\Delta E_{ST}$ with $U$ for a Type 3 junction. Within the 3NN model a minimum is observed close to $\Delta E_{ST}\sim 19$ meV at $U\sim 3.5$ eV, in reasonable agreement with (albeit larger than) the experimentally observed peak splitting. 

\clearpage
\section{Singlet-triplet excitations vs molecular size}

In the main text we have shown hat the singlet-triplet excitation energy of Type 3 junctions decreases with the size of the junctions, namely, with the length of the contacted ribbons. Here, we  present results of the interaction energy between two coupled spins for several junctions, which  corroborate this dependence. 

\begin{figure}[b!]
	\begin{center}
	\includegraphics[width=\textwidth]{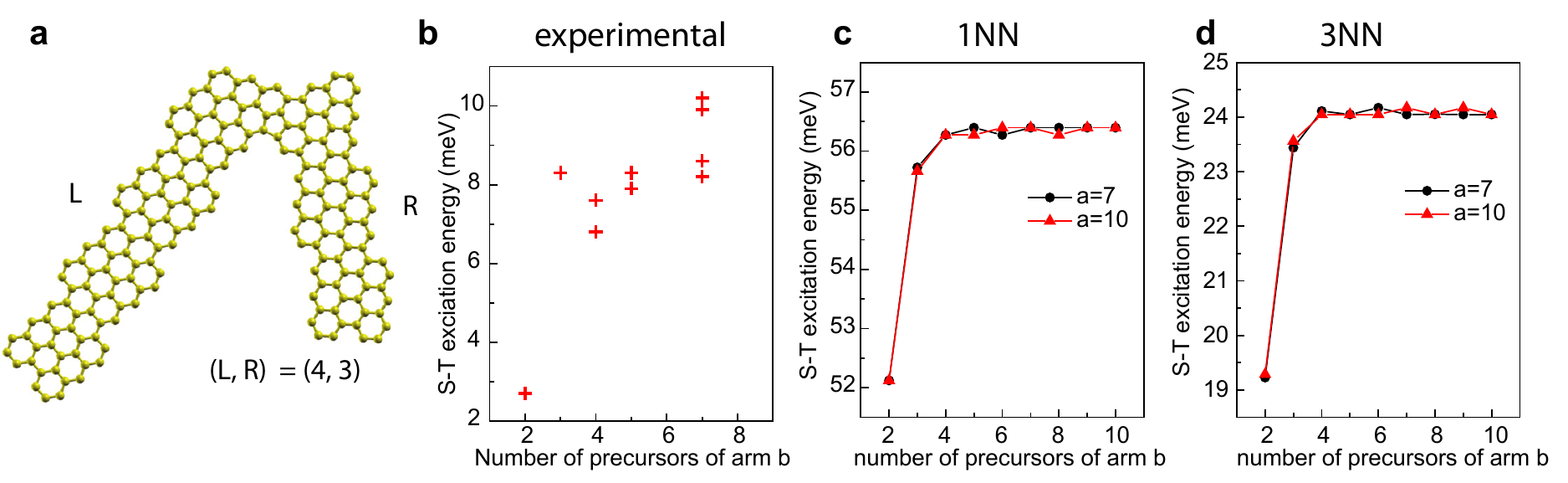}
	\end{center}
	\bfcaption{Size-dependent singlet-triplet excitation energy of Type 3 junctions.}
	{ a, Backbone structure of a Type 3 junction to illustrate  how the size of a junction is measured in terms of the number of precursor units of each of its arms (defined as arm $L$ and $R$ as in the main text).
    b, Experimentally obtained spin-spin coupling energy $J$ plotted   
    as a function of the length of arm $R$. All  junctions had  arm $L >$ 7.     
    Each data point was extracted from a fit of $dI/dV$ spectra as described in Fig.~2f. 
    c, Calculated excitation energies as a function of the length of arm $R$ within  MFH with $U=3.5$ eV in the first-nearest hopping model (1NN). Here the length of arm $a$ are fixed as 7 and 10, respectively.
    d, Same as c but for the third-nearest hopping model (3NN).  
}
\label{fig:EST}
\end{figure}

In \Figref{fig:EST}b we report experimentally extracted values of the singlet-triplet excitation energy from fits of a set of junctions, all with a similarly long arm $a$,  as a function of the length of arm $b$.
When the arm $b$ consists of two precursor units, the singlet-triplet excitation energy is relatively low (below 3 meV), but for larger length of $b$ (more than three precursor units) the energy difference raises quickly  to around 8 meV.
This trend is confirmed by the MFH simulations (\Figref{fig:EST}c,d) both with the first-nearest hopping model (1NN) and the third-nearest hopping model (3NN).
However, the experimental values are smaller than the computed ones. The origin of this could be related to the presence of a metal surface in the experiment, to the approximation of having only local Coulomb repulsion in the theory, or to limitation of the meanfield description.

\clearpage
